\documentclass[preprint,times]{elsarticle}
\pdfoutput=1

\usepackage{amsmath}  
\usepackage{amssymb}
\usepackage{subfigure}
\usepackage{ecrc}
\usepackage{rotating}
\usepackage{xcolor}
\usepackage{latexsym}
\usepackage{bm}
\usepackage{bbm}
\usepackage{floatrow}
\usepackage{breqn}
\newfloatcommand{capbtabbox}{table}[][0.45\textwidth]
\usepackage{tikz}
\newcommand\eqdef{\mathrel{\overset{\makebox[0pt]{\mbox{\normalfont\tiny\sffamily def.}}}{=}}}

\newcommand\eqij{\mathrel{\overset{\makebox[0pt]{\mbox{\normalfont\tiny\sffamily $i \neq j$}}}{=}}}

\volume{}

\jid{}
\firstpage{1}
\journalname{}
\jnltitlelogo{}
\runauth{}

\begin{document}
\begin{frontmatter}

\title{Anisotropic coalescence criterion for nanoporous materials}

\author[CEA]{V.~Gallican}
\author[CEA]{J.~Hure\corref{cor1}}

\cortext[cor1]{Corresponding author}
\address[CEA]{DEN-Service d'Etudes des Mat\'eriaux Irradi\'es, CEA, Universit\'e Paris-Saclay, F-91191, Gif-sur-Yvette, France}

\begin{abstract}
Ductile fracture through void growth and coalescence depends significantly on the plastic anisotropy of the material and on void size, as shown by experiments and/or numerical simulations through several studies. Macroscopic (homogenized) yield criteria aiming at modeling nanoporous materials have been proposed only for the growth regime, \textit{i.e.} non-interacting voids. The aim of this study is thus to provide a yield criterion for nanoporous materials relevant for the coalescence regime, \textit{i.e.} when plastic flow is localized between voids. Through homogenization and limit analysis, and accounting for interface stresses at the void-matrix interface, analytical coalescence criterion is derived under the following conditions: axisymmetric loading, orthotropic material obeying Hill's plasticity, cylindrical voids in cylindrical unit-cell. Incidentally, an orthotropic extension of the  existing isotropic modeling of interface stresses through limit analysis is described and used. The proposed coalescence criterion is then extended to account for combined tension and shear loading conditions. Numerical limit analyses have been performed under specific conditions / materials parameters to get \textcolor{black}{supposedly exact (up to numerical errors)} results of coalescence stress. A good agreement between the analytical coalescence criteria derived in this study and \textcolor{black}{numerical} results is found \textcolor{black}{for elongated spheroidal voids}, making them usable to predict the onset of void coalescence in ductile fracture modeling of nanoporous materials.
\end{abstract}


\end{frontmatter}

\section{Introduction}

Early experimental investigations have suggested \cite{henry} and shown \cite{tipper,puttick} that ductile fracture of structural materials occurs by void nucleation, growth and coalescence. The physical mechanisms and associated micromechanical models have then been described (\textit{e.g.} \cite{argon75,mcclintock,brown} for seminal contributions), while imaging techniques allow recently to assess experimentally voids evolution in porous materials in astonishing details \cite{mairewithers}. Following the pioneering works of Gurson \cite{gurson} for void growth and Thomason \cite{thomason90} for void coalescence leading to macroscopic yield criteria, homogenized models have been proposed and subsequently improved to describe porous materials at the macroscopic scale, considering the presence of voids with additional state variables, leading to ductile fracture modeling. The reader is referred to the recent reviews on this topic \cite{besson2010,benzergaleblond,pineaureview,BLNT}. 

Assuming classical continuum plasticity  (see \textit{e.g.} \cite{morinellipse} for a recent advanced model) for the behavior of the matrix surrounding the voids, most of the ductile fracture models are strictly valid only when the matrix material can be considered \textit{homogeneous} at the scale of the void. Same requirement is also necessary for models developed using crystal plasticity (see \textit{e.g.} \cite{xuhan,mbiakop,ling}), although such models allow in principle to better represent the plastic anisotropy inherent to slip systems activity for voids in single crystals  (or voids smaller than grain size in polycrystalline materials), as observed by lower-scale simulations \cite{segurado2010}. From a simple perspective\footnote{A refined perspective would be to consider additional lengthscale, \textit{e.g.} set by mobile/immobile dislocations, distances between dislocations sources.}, material homogeneity is assumed to be met for metallic materials when the characteristic length set by dislocation density (or mean free-path) $\rho^{-1/2}$ (where $\rho$ is the dislocation density) is small compared to the size of the voids, $R \gg \rho^{-1/2}$, which justify the assumption of homogeneity of the matrix material through strict scale separation. In the opposite case $R \ll \rho^{-1/2}$, only few dislocations exist at the scale of the void, and models developed assuming continuum plasticity can not rigorously be applied to predict either homogenized yield stress or void deformation under mechanical loading. In this last case, the major effect is a macroscopic hardening due to the presence of voids through pinning of dislocations (similar to Orowan's hardening with precipitates \cite{orowan}) that can be predicted and observed \cite{scattergood,lucas}, in contrast with the porosity induced softening shown by ductile fracture models. Both regimes can be observed in Discrete-Dislocation-Dynamics (DDD) simulations \cite{segurado2009,chang2015} where smallest voids start to grow after larger voids, as applied strain increases (thus dislocation density). To summarize, assuming dislocation density ranging from $10^{10}\mathrm{m^{-2}}$ (annealed state) to $10^{16}\mathrm{m^{-2}}$ (heavily work-hardened state\footnote{For such large density, dislocation sub-structures may appear, leading to \textit{heterogeneous} dislocation density associated to another lengthscale.}) for typical metallic materials \cite{fpz}, homogeneous description of the matrix surrounding the voids is met for $R\gtrsim 10\mu$m or $R\gtrsim 10$nm, respectively, keeping in mind that initial dislocation density will increase with applied strain. 
A particular case of homogeneity of the matrix surrounding the voids is however recovered for any void size at very high stress. This regime has been extensively studied through Molecular Dynamics (MD) simulations (see \cite{traiviratana,mi,brach} and references therein) where void growth comes from nucleation of dislocations from void surface \cite{lubarda}, rather than dislocations coming from the matrix to the void. One may think of this regime as a homogeneous matrix with a very high yield stress (close to the theoretical stress for dislocation nucleation) \cite{wilkerson}, and thus homogeneous description of the matrix might be also relevant in this regime. A synthesis of recent results of micro- and nano-scale voids growth through MD and DDD simulations can be found in \cite{chang2013}. 

Size effect is reported from DDD simulations \cite{segurado2009,chang2015}, \textit{i.e.} smaller voids grow slower, as a result of the limited availability of dislocations sources, inhibiting void growth by dislocations coming from the bulk. Even when plasticity takes place rather homogeneously at the scale of the void $R > \rho^{-1/2}$, another kind of size effect can be expected due to Geometrically-Necessary-Dislocations (GND) \cite{ashby}. GNDs are required to accommodate strain gradients that appear close to the voids, and control the work-hardening of the material if in excess to the Statistically-Stored-Dislocations (SSD). Such effect has been reproduced  using strain gradient plasticity constitutive equations for the matrix surrounding the voids. These strain gradient plasticity models aim at extending standard plasticity at lower scales to model the presence of GNDs by introducing additional lengthscales as described in \cite{aifantis,fleckhutchinson} for early developments (see also \cite{fleckhutchinsonwillis} and references therein for a recent synthesis of some key aspects). Void growth simulations with strain gradient plasticity \cite{fleckhutchinson2001} show a strong effect of the size of the voids on growth and strength of porous materials. Similar conclusions have been drawn using strain gradient crystal plasticity models \cite{borg2008,chaothese}. For nanoscale voids where homogeneity of the surrounding matrix may be justified, MD simulations suggest that the energy associated with the void-matrix interface (or surface energy $\gamma$) should be accounted for \cite{chang2013}. Additional stress has been shown to be required to nucleate dislocations from voids surface due to the surface energy of the interface \cite{lubarda2011}. From dimensional analysis, for $\sigma_0 R/\gamma \gg 1$ where $\sigma_0$ is the strength of the matrix, surface energy (or equivalently interface stress) is weak compared to bulk energy, and void growth is not expected to depend on surface energy, while on the contrary, the strength of the interface can make void growth size dependent. Note that this size-dependence has a different physical origin from the one described before related to dislocation density. Fig.~1 summarizes this literature review for the low stress regime, emphasizing the domain where homogenized ductile fracture type models might be relevant, and where the surface energy should be taken into account (that differentiate micro-scale from nano-scale voids). 

Homogenized models at the micro- and nano-scale have been proposed. At the micro-scale, isotropic Gurson-type void growth models have been proposed to incorporate an internal lengthscale \cite{wen2005,li2006,monchiet2013b} and showing size effects.  Similarly, isotropic Gurson type models incorporating surface stresses \cite{gurtin,monchiet2010} have been described \cite{dormieux2010,monchiet2013} and validated \cite{morin2015} accounting for interface stresses mimicking surface energy of the void-matrix interface, also exhibiting size effects. 

\begin{figure}[H]
\centering
\includegraphics[height = 7cm]{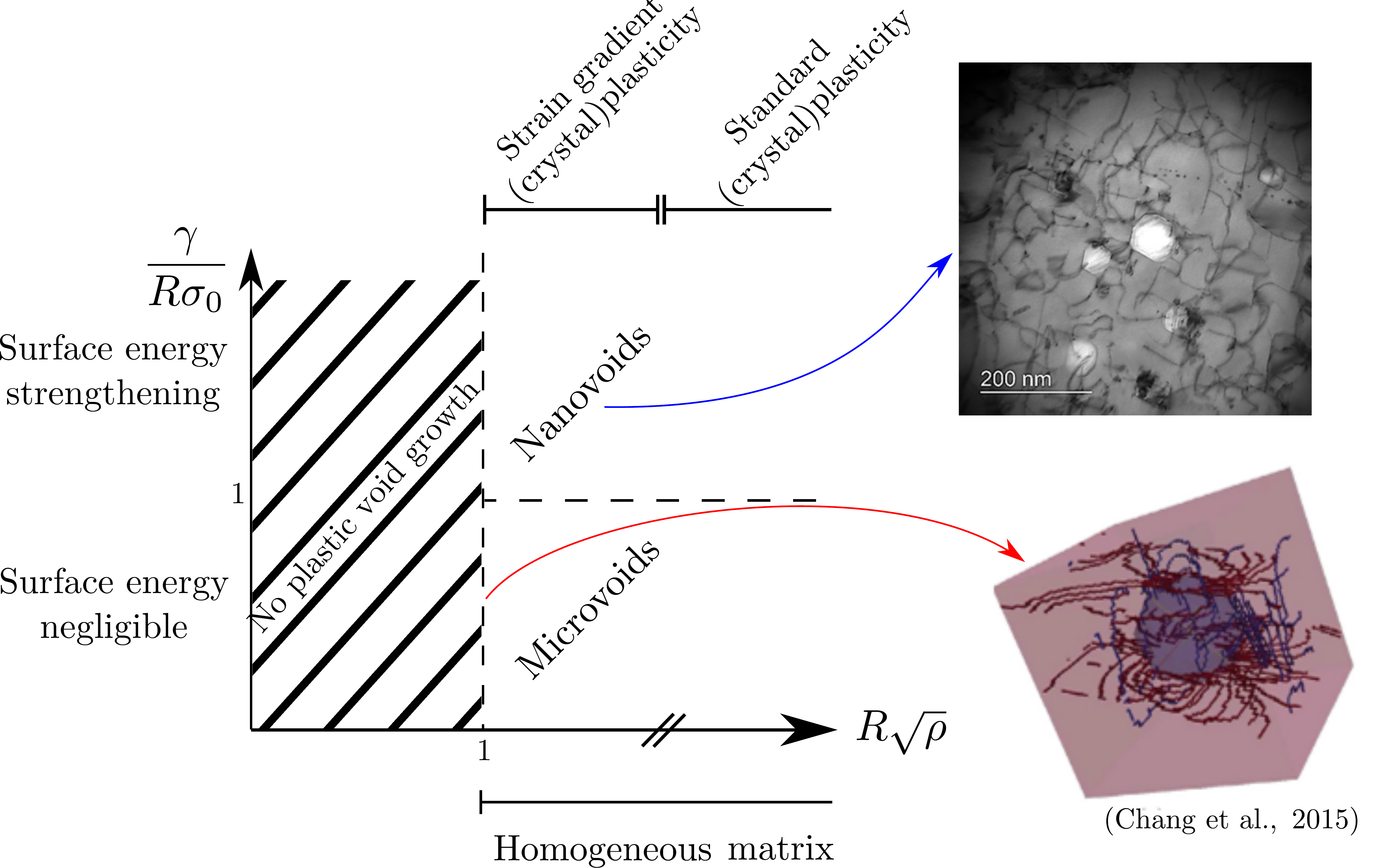}
\caption{2D map (with dimensionless parameters $R\sqrt{\rho}$ and $\gamma/\sigma_0 R$, where $R$ is the void size, $\rho$ the dislocation density, $\gamma$ the strength of the void-matrix interface, and $\sigma_0$ the strength of the matrix) used to rationalize different porous materials. Size effects (related to dislocation density and/or interface stresses) are expected for $R\sqrt{\rho} \sim 1$ and/or $\gamma/\sigma_0 R \gtrsim 1$. Image in the upper right corner is a Transmission Electron Microscope (TEM) observation of nanovoids (in white) and dislocations (black lines) in an austenitic stainless steel 304L irradiated with heavy ions (Courtesy of P.O. Barrioz). }
\label{fig1}
\end{figure}

 To the knowledge of the authors, no void coalescence models are available exhibiting size effects. This modeling of nanoporous materials is of great practical importance, for example in the nuclear field industry, where nanovoids can be formed in metallic materials as a consequence of irradiation \cite{cawthorne}. Under mechanical loading, dramatic softening of these nanoporous irradiated materials with increasing nanovoids density is observed \cite{neustroev}, and fracture surfaces exhibit nano-dimples \cite{margolin2016}.  Example of voids observed in an austenitic stainless steel 304L irradiated with heavy ions is shown on Fig.~1. The aim of this study is thus to provide a homogenized coalescence criterion for anisotropic nanoporous materials considering interface stresses. Combined with the growth criterion derived in \cite{monchiet2013}, such coalescence criterion will lead to a complete ductile failure model for nanoporous materials following the hybrid methodology that combined void growth and coalescence yield criteria described for example in \cite{benzergaleblond}.\\

The paper is organized as follows. In Section~1, the theoretical background of limit analysis and homogenization that will be used is summarized, with an emphasis on the physical relevance of the isotropic interface stresses modeling proposed in \cite{dormieux2010}. The model is in addition extended to anisotropic interface stresses, aiming at describing surface energy anisotropy observed at the nanoscale. In Section~3, theoretical coalescence stress estimates are derived based on limit analysis for axisymmetric loading conditions, and validated against numerical results from finite element simulations. The results are extended to account for combined tension and shear loading conditions in Section~4. Results are finally discussed in Section~5.

\section{Theoretical framework}

In the following, a cartesian orthonormal basis $\{\underline{e}_1,\underline{e}_2,\underline{e}_3\}$ is used with coordinates $\{x,y,z\}$, and a cylindrical orthonormal basis $\{\underline{e}_r,\underline{e}_{\theta},\underline{e}_z\}$ with coordinates $\{r,\theta,z\}$, with $\underline{e}_3 = \underline{e}_z$ (Fig.~\ref{geometry}). Underline $\underline{A}$, bold $\textbf{A}$ and double-struck $\mathbbm{A}$ symbols refer to vectors, second-order tensors and fourth-order tensors, respectively. The notation $\underline{a} \overset{S}{\otimes} \underline{b} = \underline{a} \otimes \underline{b} + \underline{b} \otimes \underline{a}$ is used, where $\otimes$ stands as the dyadic product.

\subsection{Geometry and boundary conditions}

\begin{figure}[H]
\centering
\includegraphics[height = 7cm]{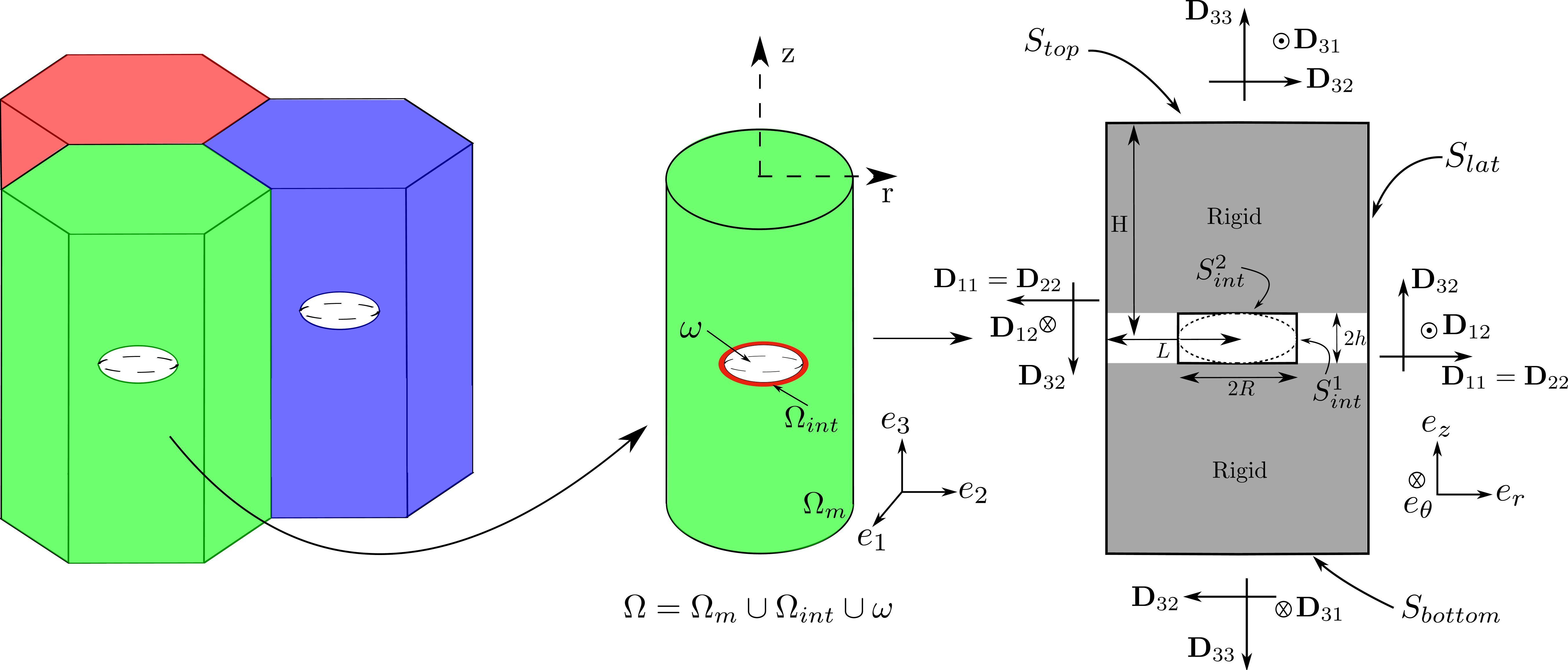}
\caption{Cylindrical unit-cell considered as an approximation of a unit-cell of a periodic array of voids of hexagonal lattice under periodic boundary conditions. Definition of geometrical parameters and loading conditions.}
\label{geometry}
\end{figure}

A cylindrical unit-cell $\Omega$ of half-height $H$ and radius $L$ containing a coaxial cylindrical void $\omega$ of half-height $h$ and radius $R$ is considered. Such geometry, which stands as an approximation of a unit-cell of a (doubly)-periodic array of hexagonal lattice under periodic boundary conditions (Fig.~\ref{geometry}), allows to derive analytical estimates of (nano)-voids coalescence in the frame of limit analysis \cite{thomasonnew2}. As detailed for example in \cite{torki} through comparisons of analytical criterion to finite-element simulations results, coalescence criterion derived for cylindrical void can a priori be used to describe spheroidal void. The effect of the unit-cell (cylindrical \textit{vs.} cubic) has been shown to be faded out considering equivalent porosities of the inter-void band \cite{torki}. Thus, homogenized coalescence criterion derived on the geometry shown in Fig.~\ref{geometry} is expected to be suitable for relevant experimental situations, \textit{i.e.} spheroidal void shapes with some equivalent porosity, keeping in mind that the relevance of homogenization (and required scale separation) might still be questionable  in the context of void coalescence (see \cite{morinunified} for a discussion about this point). 

\noindent
Three dimensionless geometrical parameters are used in the following:
\begin{equation}
W = \frac{h}{R} \ \ \ \ \ \ \ \ \ \ \chi = \frac{R}{L} \ \ \ \ \ \ \ \ \ \ c = \frac{h}{H}
\end{equation}
\noindent
where $W$ is the dimensionless void aspect ratio, $\chi$ the dimensionless length of the inter-void ligament, and $c$ the dimensionless height of the void. As coalescence corresponds to localized plastic flow in the inter-void ligament with associated elastic unloading in the regions above and below the void \cite{koplik}, boundary conditions compatible with macroscopic strain rate (see Section~\ref{la}) $\textbf{D} = D_{33} \underline{e}_3 \otimes \underline{e}_3 + D_{31}\underline{e}_3 \overset{S}{\otimes} \underline{e}_1  + D_{32}\underline{e}_3 \overset{S}{\otimes} \underline{e}_2$  for the velocity field for combined tension and shear loading conditions are (for rigid-plastic material) \cite{torki}:
\begin{equation}
\begin{aligned}
{v}_x(\underline{x})\underline{e}_1  + {v}_y(\underline{x})\underline{e}_2 &= 2Hsgn(z)\min{\left(  \frac{|z|}{h},1 \right)}\left(  D_{31} \underline{e}_1 + D_{32} \underline{e}_2 \right) &\underline{x} \in S_{lat}\\
v_z(\underline{x}) &= \pm D_{33}H &\underline{x} \in S_{top} \cup S_{bottom}\\
\end{aligned}
\label{bc}
\end{equation}
 It is worth noting that Eq.~\ref{bc} assumes that localized plastic flow height is the same as void height, which might not be relevant for flat voids $W \ll 1$, as shown in \cite{thomasonnew2,hurebarrioz}.
\subsection{Constitutive equations}

The material $\Omega_m$ is supposed to be rigid-perfectly plastic\footnote{As a classical result of limit analysis that will be used in the following that elastic strain rates vanish at limit-load.}, obeying Hill's criterion for orthotropic materials \cite{hill}, and plastic flow is associated by normality. Yield stress is denoted $\sigma_0$. Equivalent stress and strain rate are:
\begin{equation}
\left\{
\begin{aligned}
\sigma_{eq}^{H} &=  \sqrt{\frac{3}{2} \bm{\sigma}:\mathbbm{p}:\bm{\sigma}} \\
d_{eq}^{H} &= \sqrt{\frac{2}{3}\textbf{d}:\mathbbm{\hat{h}}:\textbf{d}}\\
\end{aligned}
\right.
\label{hillce}
\end{equation}
\noindent
Parameters of the fourth-order tensor $\mathbbm{\hat{h}}$ are related to those of $\mathbbm{p}$ through the relations $\mathbbm{\hat{p}} = \mathbbm{J}:\mathbbm{\hat{h}}:\mathbbm{J},\ \mathbbm{p}:\mathbbm{\hat{p}} = \mathbbm{\hat{p}}:\mathbbm{p} = \mathbbm{J}$, where $\mathbbm{J} = \mathbbm{I} - \frac{1}{3} \textbf{I} \otimes \textbf{I}$. Von Mises plasticity is recovered for specific values of tensor $\mathbbm{\hat{h}}$. Scalar anisotropy factors of the Voigt-Mandel representation of $\mathbbm{\hat{h}}$ (for details, see Appendix A), defined in previous studies on void growth \cite{benzergagld} and coalescence \cite{keralavarma}, are used in the following:

\begin{equation}
\left\{
\begin{aligned}
\hat{h}_q &= \frac{\hat{h}_{11} + \hat{h}_{22} + 4\hat{h}_{33} - 4\hat{h}_{23} - 4\hat{h}_{31} + 2\hat{h}_{12}}{6} \\
\hat{h}_{t} &= \frac{\hat{h}_{11}+\hat{h}_{22}+2\hat{h}_{66} - 2\hat{h}_{12}}{4}\\
\hat{h}_{a} &=\frac{\hat{h}_{44} + \hat{h}_{55}}{2}
\end{aligned}
\right.
\label{hqhaht}
\end{equation}
Two supplementary scalar anisotropy factors are defined:
\begin{equation}
\left\{
\begin{aligned}
\hat{h}_c &= \frac{\hat{h}_{11} + \hat{h}_{22} - 2\hat{h}_{12}}{2}\\
\hat{h}_d &= \hat{h}_{22} - \hat{h}_{11} + 2\hat{h}_{13} - 2\hat{h}_{23} 
\end{aligned}
\right.
\label{hchd}
\end{equation}
In the limit case of von Mises plasticity ($\sigma_{eq}^{H} \rightarrow \sigma_{eq}^{VM}$, $d_{eq}^{H} \rightarrow d_{eq}^{VM}$), $\{\hat{h}_q,\hat{h}_{t},\hat{h}_{a},\hat{h}_c, \hat{h}_{d}   \} \rightarrow \{1,1,1,1,0 \}$ (Appendix A).

\subsection{Limit analysis and homogenization}
\label{la}
\begin{figure}[H]
\centering
\subfigure[]{\includegraphics[height = 6cm,angle=0,origin=c]{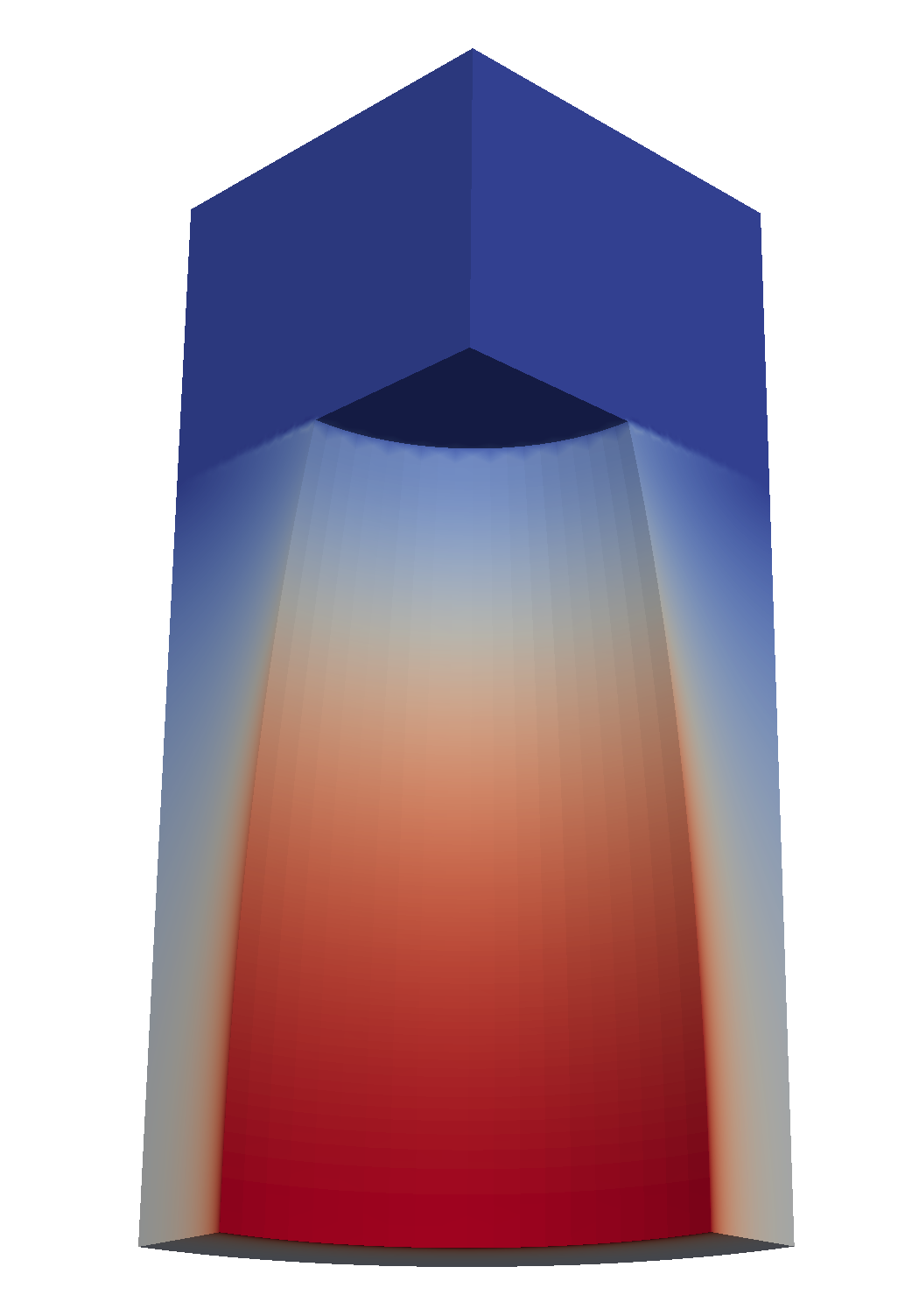}}
\hspace{0.5cm}
\subfigure[]{\includegraphics[height = 6cm,angle=0,origin=c]{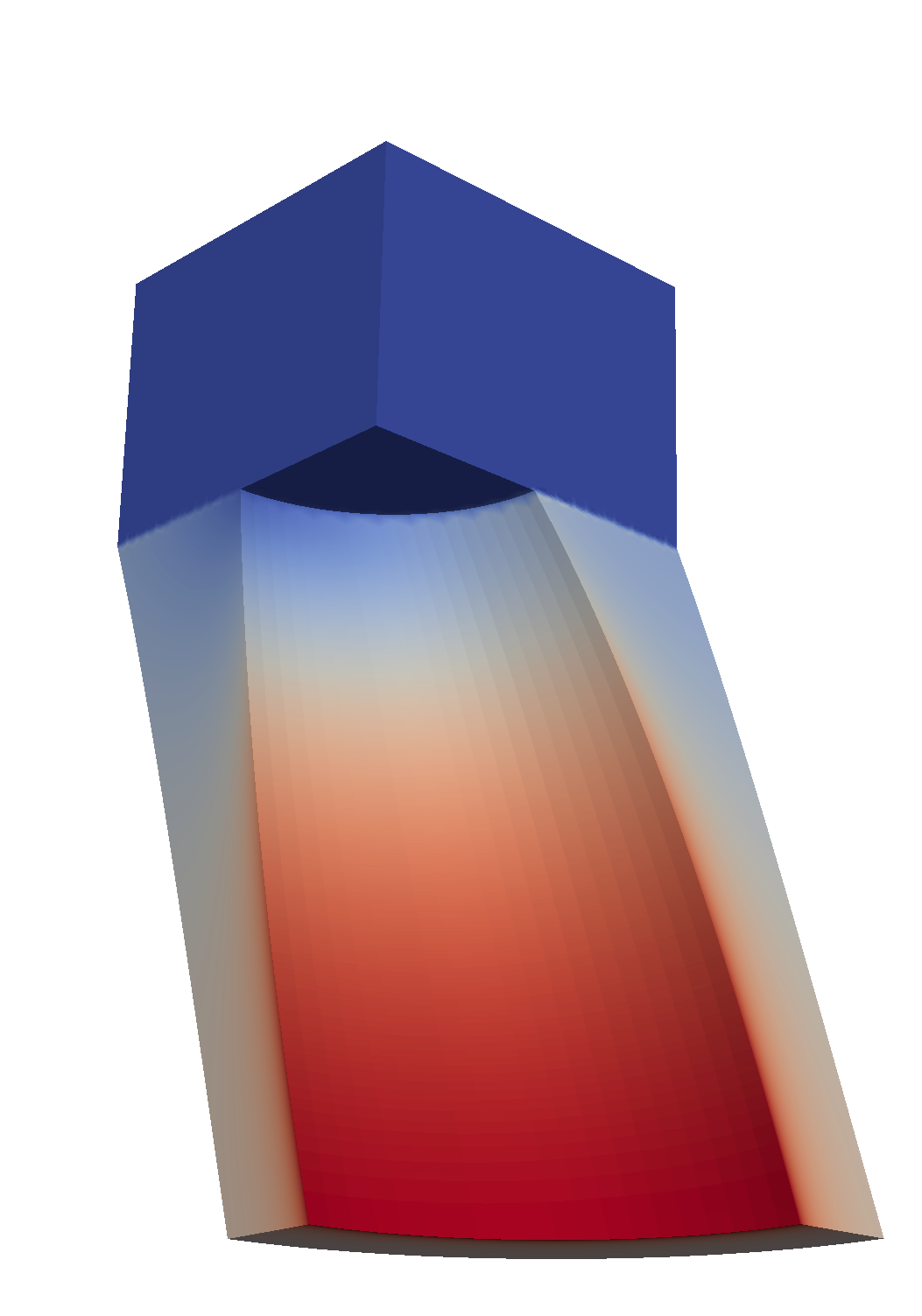}}
\vspace{-0cm}
\caption{\textcolor{black}{Visualization of the deformation of the cylindrical unit-cell with cylindrical void ($W=3$, $\chi=0.5$) resulting from the trial velocity field defined by Eqs.~\ref{ve1},\ref{sheartrial},\ref{rigidtrial} for axisymmetric (a) and combined tension and shear ($D_{31} = D_{33}$) (b) loading conditions. Colormap corresponds to the equivalent plastic strain rate (red zones correspond to the most deformed areas). Arbitrary units.}}
\label{figparaview}
\end{figure}

Limit analysis along with homogenization is used to assess the limit load of the Representative Volume Element (RVE) described in the previous section (Fig.~\ref{geometry}) with boundary conditions compatible with coalescence (Eq.~\ref{bc}). This limit load is referred to as the macroscopic yield surface of the porous cell, or to the macroscopic coalescence criterion. Upper-bound inequality of limit analysis is \cite{benzergaleblond}:
\begin{equation}
\textcolor{black}{\forall \textbf{D}\ \ \ \ \ \ \ \ \  \bm{\Sigma}:\textbf{D} \leq \Pi(\textbf{D}) }
\label{eqanalyselimite}
\end{equation}
\noindent
where $\bm{\Sigma}$ and $\textbf{D}$ corresponds to the macroscopic stress and strain rate, respectively, and are obtained through volume averaging (denoted $<.>_{\Omega}$) over the unit-cell of microscopic stress $\bm{\sigma}$ and strain rate $\textbf{d}=(\bm{\nabla} \underline{v} + \bm{\nabla}^T \underline{v})/2$. The so-called macroscopic plastic dissipation $\Pi(\textbf{D})$ can be written as \cite{benzergaleblond}:
\begin{equation}
\textcolor{black}{ \Pi(\textbf{D}) = \mathrm{inf}_{\underline{v} \in \kappa(\textbf{D})} } <\mathrm{sup}_{\bm{\sigma'}
 \in \mathcal{C}} \bm{\sigma'}:\textbf{d}  >_{\Omega}
\label{dissipation}
\end{equation}
\noindent
where $\kappa(\textbf{D})$ is the subset of velocity field compatible with $\textbf{D}$, and verifying the property of incompressibility, and $\mathcal{C}$ is the microscopic reversibility domain. Considering perfectly plastic material, obeying Hill's criterion, and plastic flow associated by normality, the macroscopic plastic dissipation is:
\begin{equation}
\textcolor{black}{ \Pi(\textbf{D}) = \mathrm{inf}_{\underline{v} \in \kappa(\textbf{D})} } <\sigma_{eq}^H d_{eq}^H  >_{\Omega}
\label{dissipationmises}
\end{equation}
\noindent
For brevity, the term $<\sigma_{eq} d_{eq}  >_{\Omega}$ computed for a given velocity field will be referred to as the macroscopic plastic dissipation, although only an upper-bound of it. Finally, for differentiable plastic dissipation, the inequality of limit analysis (Eq.~\ref{eqanalyselimite})) is equivalent to:
\begin{equation}
\bm{\Sigma} = \frac{\partial \Pi(\textbf{D})}{\partial \textbf{D}}
\label{criterion}
\end{equation}
\noindent
 which gives the macroscopic yield surface of the porous cell, also called macroscopic coalescence criterion. Limit analysis relies on the choice of a \textit{trial} velocity field to compute the plastic dissipation. The continuous velocity field proposed by Keralavarma and Chockalingham \cite{keralavarma} is used: in the inter-void ligament $\{|z|\leq h; R \leq r \leq L \}$  
\begin{equation}
\left\{
\begin{aligned}
v_r^{KC}(r,z) & = \frac{3HD_{33}}{4h}\left(1-\frac{z^2}{h^2}\right)\left(\frac{L^2}{r}-r    \right) \\
v_z^{KC}(r,z) &= \frac{3HD_{33}}{2h}\left(z-\frac{z^{3}}{3h^2} \right)\\
\end{aligned}
\right.
\label{ve1}
\end{equation}
\noindent
In addition, some remarks will be made in the following regarding the use of alternative trial velocity fields proposed in \cite{thomasonnew2}. The velocity field of Eq.~\ref{ve1} is compatible with $\textbf{D}$ (through Eq.~\ref{bc}) only in axisymmetric loading conditions $\textbf{D} = D_{33} \underline{e}_3 \otimes \underline{e}_3$. In the presence of shear, it shall be supplemented by a shear trial velocity field in the intervoid ligament \cite{torki}:
\begin{equation}
\left\{
\begin{aligned}
\textcolor{black}{v_r^{shear}(r,\theta,z)} &  \textcolor{black}{=\frac{2zH}{h}(D_{31}\cos{\theta} + D_{32}\sin{\theta}    )} \\
\textcolor{black}{v_{\theta}^{shear}(r,\theta,z)} & \textcolor{black}{=\frac{2zH}{h}(-D_{31}\sin{\theta} + D_{32}\cos{\theta}    )}\\
\end{aligned}
\right.
\label{sheartrial}
\end{equation}
\noindent
The upper and lower parts of the unit-cell are rigid: 
\begin{equation}
\begin{aligned}
\underline{v}(r,z\geq |h|) &= sgn(z)\,H\left(  2D_{31} \underline{e}_1 + 2D_{32} \underline{e}_2 + D_{33} \underline{e}_3 \right) \\
\end{aligned}
\label{rigidtrial}
\end{equation}
This velocity field \textcolor{black}{(Fig.~\ref{figparaview})} has been shown to lead to good estimates of coalescence stress for both isotropic (von Mises) and anisotropic (Hill) \cite{keralavarma} matrix yield criterion for $W \gtrsim 1$. For flat voids $W \ll 1$, it does not lead to good estimates of coalescence stress (unless phenomenological modifications are used) and refined velocity fields are required \cite{hurebarrioz}.

\subsection{Interface stresses}

In line with the theory of Gurtin and Murdoch of surface stresses in solids \cite{gurtin}, Kondo and Dormieux \cite{dormieux2010} proposed to model nanoporous ductile isotropic materials by considering that a (small) layer at the void-matrix interface (denoted $\Omega_{int}$ on Fig.~\ref{geometry}) has a different plastic mechanical behavior from the matrix. Using limit analysis and homogenization, Gurson model was in this way extended to nanoporous material, showing size effects through a parameter that relates the (3D) yield stress of the matrix to the (2D) yield stress of the interface. The homogenized model was then extended to spheroidal voids in isotropic matrix \cite{monchiet2013}, and the numerical integration of the set of constitutive equations has been described \cite{morin2015}.
The plastic dissipation corresponding to this model is:
\begin{equation}
\Pi(\textbf{D}) = \mathrm{inf}_{\underline{v} \in \kappa(\textbf{D})}  <\sigma_{eq}^{VM} d_{eq}^{VM}  >_{\Omega} =  \mathrm{inf}_{\underline{v} \in \kappa(\textbf{D})}  \frac{1}{vol(\Omega)}\left[  \int_{\Omega_m} \sigma_0 d_{eq}^{VM}\,dV +  \int_{\Omega_{int}} \sigma_0^{interface} d_{eq}^{VM}\,dV \right] 
\label{dissipationmisesnano}
\end{equation}
\noindent
where $\Omega_{int}$ is the volume of the small layer at the interface between the void and the matrix $\Omega_m$. As proposed in \cite{dormieux2010} and derived through Taylor expansion in \cite{monchiet2010}, the second integral of Eq.~\ref{dissipationmisesnano} can be reduced to a 2D integral on the interface, by assuming that the thickness of the layer is small compared to other dimensions:
\begin{equation}
\Pi(\textbf{D}) = \mathrm{inf}_{\underline{v} \in \kappa(\textbf{D})}  <\sigma_{eq}^{VM} d_{eq}^{VM}  >_{\Omega} =  \mathrm{inf}_{\underline{v} \in \kappa(\textbf{D})}  \frac{1}{vol(\Omega)}\left[  \int_{\Omega_m} \sigma_0 d_{eq}^{VM}\,dV +  \int_{S_{int}} k_{int} d_{S,eq}^{VM}\,dS \right] 
\label{dissipationmisesnano2}
\end{equation}
\noindent
where \textcolor{black}{$k_{int} = h\sigma_0^{interface}$} is the 2D yield stress of the interface, and $d_{S,eq}^{VM}$ the 2D equivalent strain rate of the interface:
\begin{equation}
d_{S,eq}^{VM} = \sqrt{\frac{2}{3}\left[\textbf{d}^S:\textbf{d}^S + (\mathrm{tr}\textbf{d}^S)^2      \right]}
\label{dseqar1}
\end{equation}
\noindent
$\textbf{d}^S$ is the 2D strain rate of the interface, obtained by projecting the 3D strain rate  $\textbf{d}$ to the tangent plane of the interface through the projector tensor $\textbf{P}(\underline{n}) = \textbf{I} - \underline{n}\otimes \underline{n}$, where $\underline{n}$ is the unit normal to the interface \cite{monchiet2013}:
\begin{equation}
\textbf{P}(\underline{n}).\textbf{d}.\textbf{P}(\underline{n}) = \left( \begin{array}{ccc}
0 & 0 & 0 \\
0 & &  \\
0 & \multicolumn{2}{c}{\smash{\raisebox{.5\normalbaselineskip}{[$\textbf{d}^S$]}}}
  \end{array} \right)
\end{equation}
in the local orthonormal frame where the first axis is $\underline{n}$. \textcolor{black}{The yield criterion of the 2D interface associated with Eq.~\ref{dseqar1} is the plane stress version of the 3D yield criterion defined in Eq.~\ref{hillce}: $[3/2]\bm{\sigma}_{2D}:\bm{\sigma}_{2D}- k_{int}^2\leq 0$ \cite{morinthese}.} A dimensionless parameter $\Gamma$ is defined to relate the (3D) yield stress of the matrix to the (2D) yield stress of the interface:
\begin{equation}
\Gamma = \frac{k_{int}}{\sigma_0 R}
\label{gamma}
\end{equation}
For fixed mechanical properties of the interface and matrix ($k_{int}$ and $\sigma_0$), the lower the value of $R$, the higher the value of $\Gamma$, thus the higher the plastic dissipation (from Eq.~\ref{dissipationmisesnano2}) and the higher the homogenized stress. The parameter $\Gamma$ is therefore a way to reproduce size effects. Eq.~\ref{dissipationmisesnano2} can be recovered by noting that, in $\Omega_{int}$, the strain rate is equal to \cite{monchiet2010}:
\begin{equation}
\textbf{d}^{\star} = \left( \begin{array}{ccc}
-\mathrm{tr}(\textbf{d}^S) & 0 & 0 \\
0 & &  \\
0 & \multicolumn{2}{c}{\smash{\raisebox{.5\normalbaselineskip}{[$\textbf{d}^S$]}}}
  \end{array} \right)
\label{dstar}
\end{equation}
as the out-of-plane shear strain rates become negligible compared to other terms when the thickness of the layer goes to zero, and normal strain rate ensures incompressibility. Thus, plastic dissipation of the interface can be computed as:
\begin{equation}
\begin{aligned}
\int_{\Omega_{int}} \sigma_0^{interface} d_{eq}^{VM}\,dV  &= \int_{\Omega_{int}} \sigma_0^{interface} d_{eq}^{\star,VM}\,dV   \\
&= \int_{S_{int}} h \sigma_0^{interface} \sqrt{\frac{2}{3}\textbf{d}^{\star}:\textbf{d}^{\star}}\,dS \\
&= \int_{S_{int}} k_{int} d_{S,eq}^{VM} \,dS 
\end{aligned}
\label{deriveq2D}
\end{equation}
\noindent 
with $k_{int} = h\sigma_0^{interface}$, which corresponds to Eq.~\ref{dissipationmisesnano2}. The physical interpretation of the new term in Eq.~\ref{dissipationmisesnano2} is an energetic cost associated to the deformation of the void-matrix interface. For the case of a diagonal 2D strain rate tensor $\textbf{d}^{S} = \alpha \textbf{1}$, the integral reduces to $k_{int}|\dot{\mathcal{S}}|$ (using the relation $\dot{\mathcal{S}}/\mathcal{S}=\mathrm{tr}\textbf{d}^{S}$), where $\mathcal{S}$ is the surface of the interface. This surface energy is quantitatively consistent with the modeling of surface tension $\gamma \mathcal{S}$ in the case
 of an increase of the surface of the interface, with $k_{int} = \gamma$ of the order of $1\mathrm{J.m^{-2}}$ for surface tension of solids/fluids interfaces \cite{crystallium1}. For the case of general 2D strain rate tensor $\textbf{d}^{S}$, the integral reduces to $a k_{int}|\dot{\mathcal{S}}|$, where $a$ is a prefactor\footnote{$a = \sqrt{\frac{2}{3}\left[1+ \frac{1+\alpha^2 + 2\beta^2}{(1+\alpha)^2}     \right]}$, with $d_{22} = \alpha d_{11}$ and $d_{12} = \beta d_{11}$.} depending on the relative magnitude of the component of $\textbf{d}^{S}$, thus such modeling reproduces only \textcolor{black}{qualitatively} (or quantitatively but through an effective surface energy $k_{int} = \gamma/a$) the surface energy associated with the increase of the surface of the interface. In \cite{monchiet2013}, another term was proposed in Eq.~\ref{dissipationmisesnano2} equal to $k_{res} \mathrm{tr}\textbf{d}^{S}$ and related to residual stresses in the theory of surface stresses \cite{gurtin}. This term has the advantage of modeling quantitatively the energetic cost associated with a surface energy in any situation (including tension and compression as shown in \cite{monchiet2013}). However, for practical situations in ductile fracture, \textit{i.e.} large positive stress triaxialities, both terms lead to similar yield criteria. Therefore, in the following, only plastic dissipation of Eq.~\ref{dissipationmisesnano2} is considered, keeping in mind that such modeling of nanoporous material by accounting for surface energy is valid only in tension, and for situations where the interface strain (rate) tensor is close to identity.

From a qualitative point of view, such simple modeling may also be relevant to describe other kind of interface dissipation, like for example the one related to GNDs, through an effective value of surface energy. This point will be further described in Section~\ref{conclusion}.

\begin{figure}[H]
\captionsetup[subfigure]{skip=-100pt}
\centering
\subfigure{\includegraphics[height = 7cm,angle=270,origin=c]{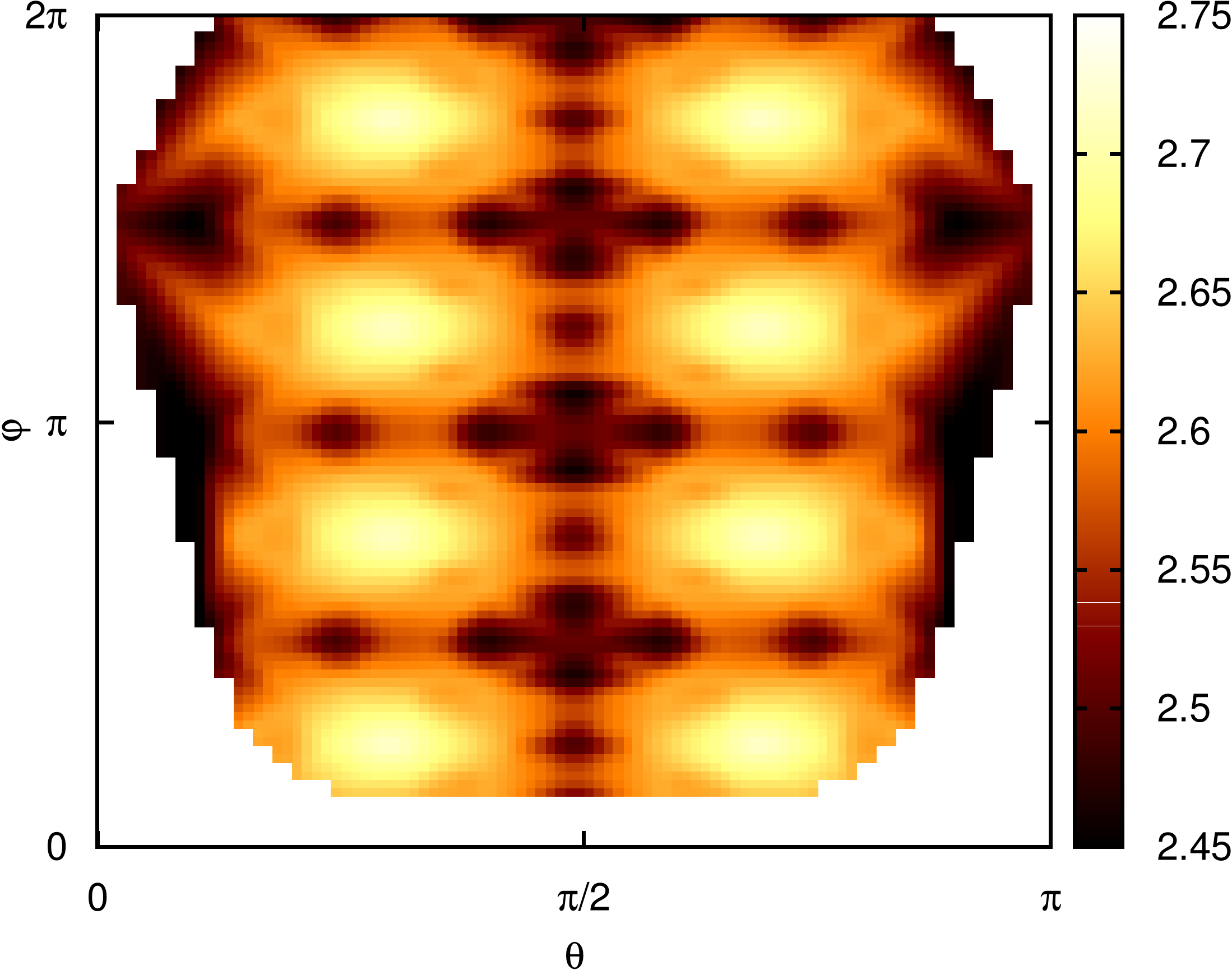}}
\hspace{0.5cm}
\subfigure{\includegraphics[height = 7cm,angle=270,origin=c]{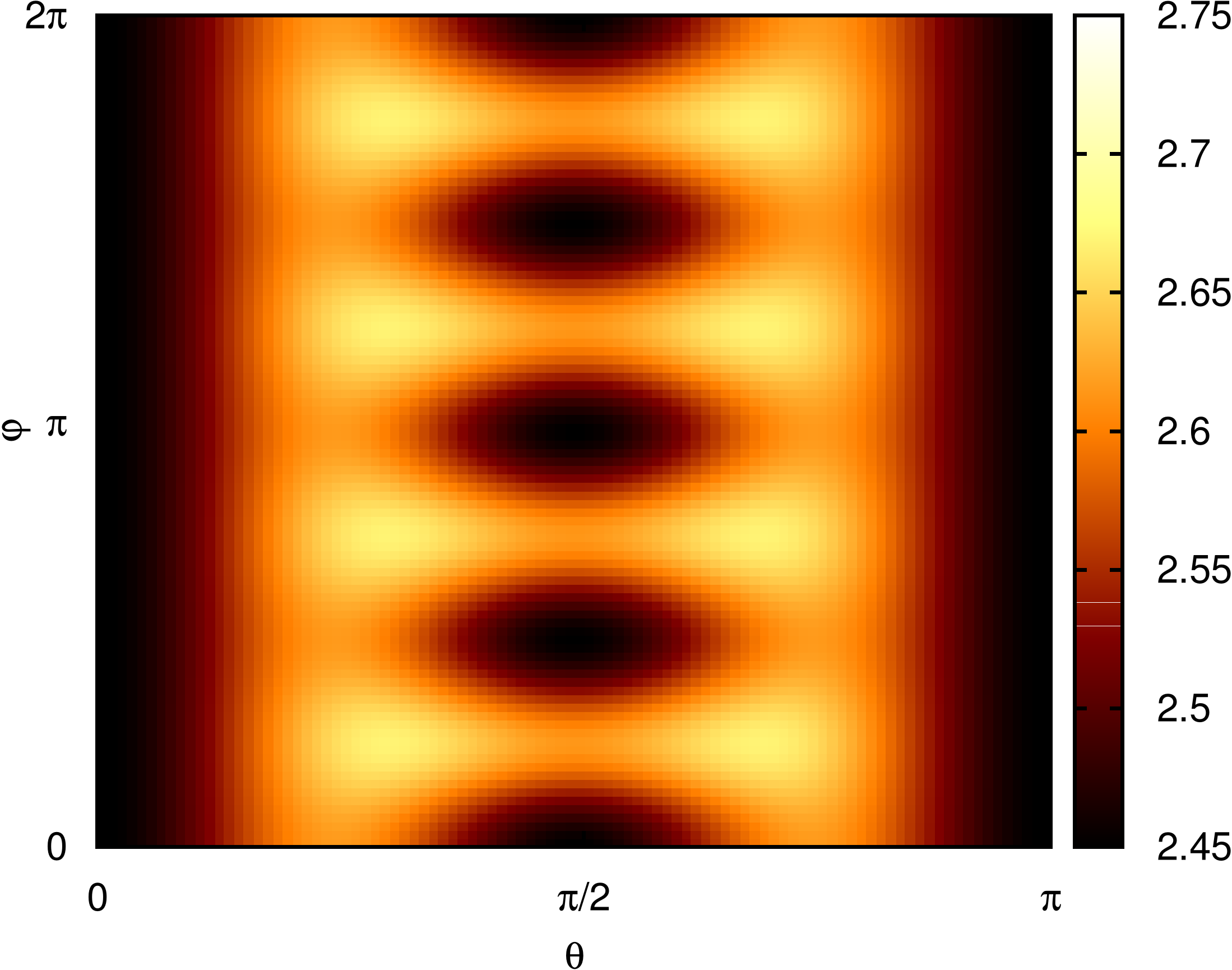}}
\vspace{-1cm}
\caption{Representation of the anisotropy of surface energy $\gamma(\underline{n})$ (in $\mathrm{J.m^{-2}}$) of pure Iron in spherical coordinates (with $\theta$ and $\phi$ the coordinates of the normal $\underline{n}$: (a) experimental values extrapolated linearly from data given in \cite{crystallium1,crystallium2}, (b) numerical values from Eq.~\ref{defgamman} where the parameters of $\mathbbm{\hat{h}}$ have been adjusted \textcolor{black}{($\mathbbm{\hat{h}}_{11}=\mathbbm{\hat{h}}_{22}=\mathbbm{\hat{h}}_{33}=1$; $\mathbbm{\hat{h}}_{44}=\mathbbm{\hat{h}}_{55}=\mathbbm{\hat{h}}_{66}=0.59$; $\mathbbm{\hat{h}}_{12}=\mathbbm{\hat{h}}_{13}=\mathbbm{\hat{h}}_{23}=0$, with $k_{int}=\sqrt{6}\mathrm{J.m^{-2}}$)}.}
\label{figgamma}
\end{figure}

This model is still limited to the isotropic case, where surface energy does not depend on the normal to the interface. However, surface energy is known to be dependent on the orientation of the interface, leading to crystal and nano-voids shapes \cite{crystallium2}. Thus, an extension of the model described in \cite{dormieux2010,monchiet2013} and summarized by Eq.~\ref{dissipationmisesnano2} is proposed by considering Hill criterion instead of von Mises criterion for the plasticity of the matrix and of the interface. The plastic dissipation is defined as:
\begin{equation}
\Pi(\textbf{D}) = \mathrm{inf}_{\underline{v} \in \kappa(\textbf{D})}  <\sigma_{eq}^{H} d_{eq}^{H}  >_{\Omega} =  \mathrm{inf}_{\underline{v} \in \kappa(\textbf{D})}  \frac{1}{vol(\Omega)}\left[  \int_{\Omega_m} \sigma_0 d_{eq}^{H}\,dV +  \int_{S_{int}} k_{int} d_{S,eq}^{H} \,dS\right] 
\label{dissipationmisesnano3}
\end{equation}
\noindent
where the equivalent Hill strain rate for the interface, following the derivation of Eq.~\ref{deriveq2D}, is given by:
\begin{equation}
d_{S,eq}^{H} = \sqrt{\frac{2}{3}\textbf{d}^{\star}_{orth}:\mathbbm{\hat{h}}:\textbf{d}^{\star}_{orth}}
\end{equation}
where $\textbf{d}^{\star}_{orth}$ is the interface strain rate tensor written in the same orthonormal basis as $\mathbbm{\hat{h}}$. Eq.~\ref{dissipationmisesnano3} accounts for both anisotropy of the matrix and of the interface through the parameters of $\mathbbm{\hat{h}}$ (that may be different for both). As an example, for the case $\textbf{d}^{S} = \alpha \textbf{1}$, the integrand can be written as:
\begin{equation}
\begin{aligned}
 k_{int} d_{S,eq}^{H} d\mathcal{S} &= k_{int} \sqrt{\frac{2}{3}\textbf{d}^{\star}_{orth}:\mathbbm{\hat{h}}:\textbf{d}^{\star}_{orth}} d\mathcal{S} \\
  &= \alpha k_{int}  \sqrt{\frac{2}{3}\left[^T \textbf{R}(\underline{n}).\textbf{d}^{\star}.\textbf{R}(\underline{n})\right]:\mathbbm{\hat{h}}:\left[ ^T\textbf{R}(\underline{n}).\textbf{d}^{\star}.\textbf{R}(\underline{n})\right]}\, d\mathcal{S} \\
 &\eqdef \gamma(\underline{n})\, d\mathcal{\dot{S}}
\end{aligned}
\label{defgamman}
\end{equation}  
\noindent
where $\underline{n}$ is the unit normal to the interface and $\textbf{R}(\underline{n})$ the associated rotation matrix that goes from the local frame of the interface to the global frame in which the anisotropy tensor $\mathbbm{\hat{h}}$ is defined. Eq.~\ref{defgamman} makes clear that choosing the parameters of $\mathbbm{\hat{h}}$ is a way to reproduce the anisotropy of surface energy $\gamma(\underline{n})$. As an illustrative example, the experimental anisotropy of surface tension of pure Iron \cite{crystallium1,crystallium2} is compared to the model of Eq.~\ref{defgamman} where the parameters of $\mathbbm{\hat{h}}$ have been adjusted in Fig.~\ref{figgamma}. Overall good agreement indicates that the anisotropic extension of the model proposed in \cite{dormieux2010} may be an effective way to represent the anisotropy of surface energy, or a way to account for anisotropic interfacial energy.

\subsection{\textcolor{black}{Numerical simulations}}
\label{numana}
\subsubsection{\textcolor{black}{Numerical limit analysis}}
Limit analysis relies on finding a \textit{trial velocity field} to get an upper-bound of the plastic dissipation from Eq.~\ref{dissipationmises}, leading to an (analytical) upper-bound of the homogenized yield criterion from Eq.~\ref{criterion}. How far the upper-bound (or the estimate if approximations that are not upper-bound preserving are used to compute the plastic dissipation) is from the true yield criterion needs to be assessed. This requires finding the velocity field solution of the microscopic boundary value problem (BVP) (Fig.~\ref{geometry}, with boundary conditions Eq.~\ref{bc} and constitutive equations Eq.~\ref{hillce}) for which the infimum of the right-hand side of Eq.~\ref{dissipationmises} is obtained. As shown in \cite{madou}, this can be done by performing a small strain elastoplastic finite-element simulation, with no geometry update, perfect plasticity with implicit integration framework and one large loading step (such as elastic strains are negligible compared to plastic strains). Under these assumptions, limit analysis is equivalent to the BVP, where the displacement field obtained in finite-element simulation stands as the velocity field. Plastic dissipation can be computed (Eq.~\ref{dissipationmises}), or macroscopic stress (Eq.~\ref{criterion}) through volume averaging. Another way of finding the macroscopic stress or homogenized yield criterion is to perform a small strain elastoplastic finite-element simulation, with no geometry update, perfect plasticity, but with multiple loading steps until saturation of the macroscopic stress (obtained again through volume averaging), as for example shown in \cite{tekoglushear}. Both methods lead to the same values of macroscopic stresses, and thus homogenized yield criterion. In practice, the one loading step method requires a large number of iterations of the Newton-Raphson algorithm to get the equilibrium, while the multiple loading steps method may require a large number of steps, but with only few iterations of the Newton-Raphson algorithm at each step.

Numerical simulations are performed in this study with the finite-element software Cast3M \cite{castem} with no geometry update, elastic-perfectly plastic material (with Young's modulus $E$, Poisson's ratio $\nu$ and yield stress $\sigma_0$). Young's modulus and Poisson's ratio are set to $10^4 \sigma_0$ and $0.49$, respectively, but limit-loads do not depend on these parameters as a classical result of limit analysis. Multiple loading steps method is used. Depending on the loading conditions (axisymmetric or tension/shear), either 2D or 3D meshes are used, with fully integrated quadratic elements. For combined tension and shear loading conditions, pseudo-periodic boundary conditions are used on the lateral surface of the cylindrical unit-cell: $\forall z\ \underline{v}(x,y,z) = \underline{C}^{ste}(z)$. \textcolor{black}{The boundary conditions used in this study are the same as the ones used recently in \cite{torki2017}.} Interface stresses are accounted for in two different ways. First method is to add shell elements at the interface between the matrix and the void, with yield stress $\sigma_0^S$. The thickness of the shell $t$ is set such as $k_{int} = \sigma_0^S t$. For small enough values of the thickness $t$ (assessed numerically), bending moments of the shell are negligible compared to stretching, and this finite-element modeling is consistent with Eq.~\ref{deriveq2D}. Second method is to add one (thin) layer of volumic elements. While being less efficient numerically, this second method allows straightforward implementation of anisotropic interface stresses than the first one. In any case, both methods lead to the same values of limit loads, and mesh convergence has been checked for all simulations. \textcolor{black}{Finally, localization of plastic flow in the inter-void ligament (and almost rigid motion of the domains below and above the void) have been checked for all numerical results presented hereafter, confirming that the limit-loads obtained correspond to void coalescence.}

\begin{figure}[H]
\centering
\includegraphics[height = 5cm]{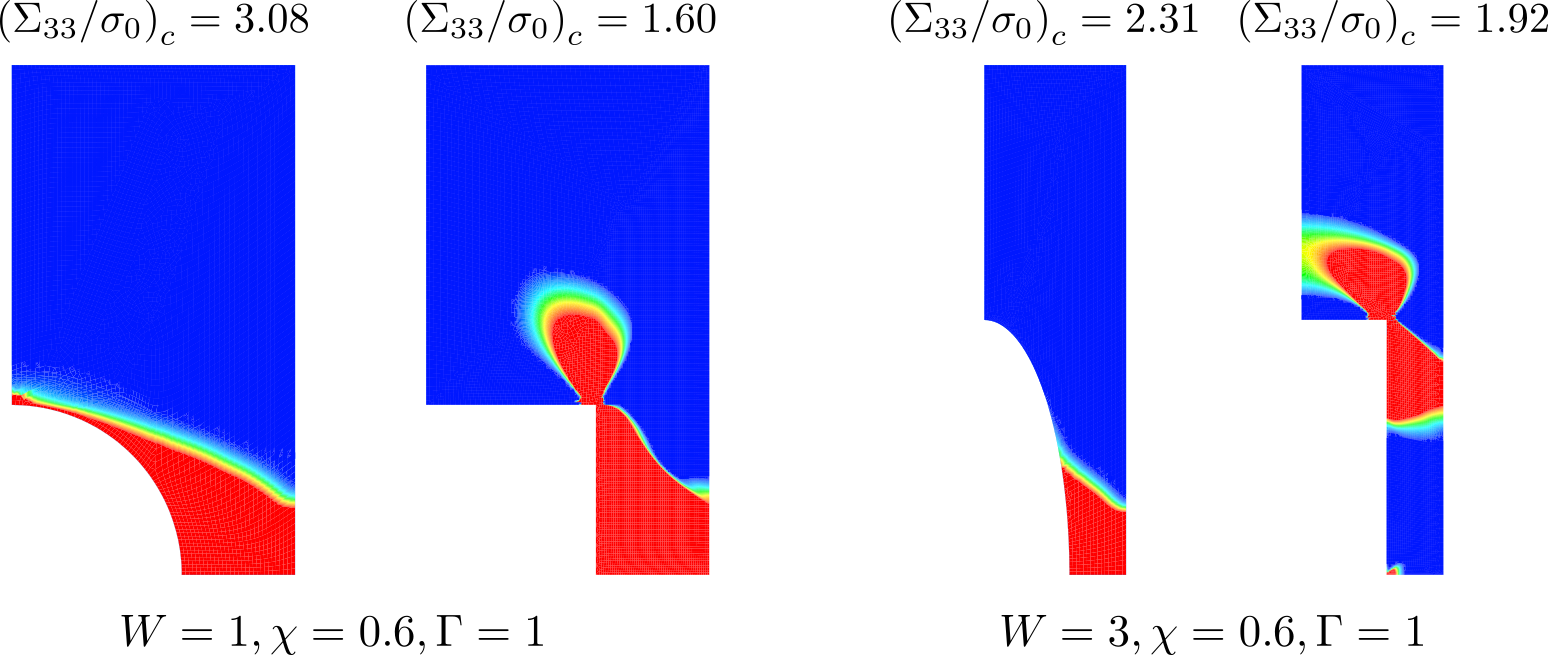}
\caption{\textcolor{black}{Numerical limit analysis of spheroidal and cylindrical voids in axisymmetric loading conditions $\textbf{D} = D_{33} \underline{e}_3 \otimes \underline{e}_3$. Equivalent strain (rate) at limit load in arbitrary units (red zones correspond to most deformed areas).}}
\label{figstrainfield}
\end{figure}

In absence of interface stresses, it has been shown on particular cases in \cite{thomasonnew2,hurebarrioz} through numerical limit analysis that void shape (cylindrical \textit{vs.} spheroidal) has only a weak influence on coalescence limit loads. This justified to perform numerical limit analysis only considering cylindrical voids to validate analytical limit loads estimates (obtained considering also cylindrical voids), knowing that these estimates could be used in principle for practical situations, \textit{i.e.} spheroidal voids. However, accounting for interface stresses, void shape has a strong influence on both coalescence limit loads and plastic strain rate field at limit load, as can be seen on Fig.~\ref{figstrainfield}, where plastic flow does not localize in the inter-void ligament for cylindrical voids.

Therefore, numerical limit analysis is performed considering only spheroidal voids, relevant for applications, and results are compared to the analytical estimates obtained through analytical limit analysis considering cylindrical voids (as trial velocity fields are easier to get for such geometry). This implies that, although upper-bound inequality of limit analysis is used (Eq.~\ref{eqanalyselimite}), analytical estimates of coalescence stress might not be upper-bound of numerical coalescence stress for spheroidal voids. However, these analytical estimates are expected to be good estimates for coalescence stress of spheroidal voids.

\subsubsection{\textcolor{black}{Parameters range}}
Two sets of material parameters have been chosen to perform the simulations. First one corresponds to a von Mises material, second one to a transversely isotropic material (with respect to $\underline{e}_3$ axis, see Fig.~\ref{geometry}) already used in \cite{kera2011} that shows a strong effect of anisotropy on coalescence stress. Parameters of the Voigt-Mandel (Appendix A) representation of fourth-order tensor $\mathbbm{\hat{h}}$ (Eq.~\ref{hillce}) for these two materials are given in Tab.~1.

\begin{table}[H]
\begin{tabular}{c||c|c|c|c|c|c}
Material reference & $\hat{h}_{11}$ & $\hat{h}_{22}$ & $\hat{h}_{33}$ & $\hat{h}_{44}$ & $\hat{h}_{55}$ & $\hat{h}_{66}$ \\
\hline
\hline
Isotropic & 1.0 & 1.0 & 1.0 & 1.0 & 1.0 & 1.0\\
\hline
Anisotropic & 3/7 & 3/7 & 9/7 & 1.0 & 1.0 & 3/7\\ 
\end{tabular}
\label{tab1}
\caption{Parameters of the Voigt-Mandel representation of fourth-order tensor $\mathbbm{\hat{h}}$ (Eq.~\ref{hillce}) used to perform numerical simulations, other terms being equal to zero. Anisotropic material corresponds to \textit{Material (iv)} of \cite{kera2011}.}
\end{table}

\textcolor{black}{The values of the dimensionless parameter $\Gamma$ considered in this study have been chosen as follows. On one hand, assuming classical continuum plasticity to hold at the scale of a void of typical size $R$ requires $R\gtrsim 1/\sqrt{\rho}$, where $\rho$ is the dislocation density. Heavily deformed material can have a dislocation density up to $10^{16}\,\mathrm{m^{-2}}$, thus the model proposed is restricted to void size larger than 10nm. On the other hand, the yield stress of the matrix can be estimated through Taylor-like hardening equation $\sigma_0 \sim \mu b \sqrt{\rho}$, where $\mu$ is the shear modulus, and $b$ the Burgers' vector. The maximal value of $\Gamma$ can thus be written as $\Gamma = k_{int}/[\sigma_0 R] \lesssim k_{int}/[\mu\,b] < 1$, where values of $\mu = 65$GPa, $b=0.2$nm \cite{phdHan} and $k_{int}=2\mathrm{J.m^{-2}}$ (Fig.~\ref{figgamma}) for austenitic stainless steel have been used. Therefore, $\Gamma =0, 0.5, 1$ values have been used hereafter.}

\textcolor{black}{The dependence of the coalescence stress on void aspect ratio $W$ should be assessed for values ranging from well above to well below $1$, in order for the criterion to be applicable for situations ranging from large (positive) stress triaxialities (where void growth will lead to elongated voids before coalescence) to intense shearing (where void rotation can lead to penny-shaped cracks). \textcolor{black}{However, preliminary numerical limit analysis simulations have shown that no localization of plastic flow in the inter-void ligament occurs under axisymmetric loading conditions for flat voids ($W \ll 1$) in presence of interface stresses ($\Gamma = 0.5, 1$). Although such observation does not imply that coalescence of voids with interface stresses would not happen under specific conditions, \textit{e.g.}, intense shearing, the case of flat voids is not considered in this study which focuses on elongated voids.} Thus, the range of void aspect ratio used in the following is $W \in [1:3]$. The effect of the dimensionless length of the inter-void ligament has been assessed in the range $\chi \in [0.3:0.7]$.}

\section{Axisymmetric loading conditions}

Axisymmetric loading conditions $D_{ij} \eqij 0$ are considered in this section. As a result of the definition of plastic dissipation (Eq.~\ref{dissipationmisesnano3}) and the definition of the yield criterion through limit analysis (Eq.~\ref{criterion}, reducing to $\Sigma_{33} = \Pi/D_{33}$ for axisymmetric loading conditions as $D_{11} = D_{22} = 0$), the coalescence criterion can be written as:
\begin{equation}
\left(  \frac{\Sigma_{33}}{\sigma_0} \right)_c \leq \left(  \frac{\Sigma_{33}}{\sigma_0} \right)_{matrix} + \left(  \frac{\Sigma_{33}}{\sigma_0} \right)_{interface}
\label{sumtwoterms}
\end{equation}
The first term on the right-hand side of Eq.~\ref{sumtwoterms} comes from the plastic dissipation in the matrix, while the second term comes from the dissipation on the interface. First term has already been computed for the velocity field considered in this study in \cite{keralavarma}:
\begin{equation}
\left(   \frac{\Sigma_{33}}{\sigma_0}\right)_{matrix} \approx \sqrt{\frac{6}{5} \hat{h}_q} \left[b \ln\frac{1}{\chi^2} + \sqrt{b^2+1} - \sqrt{b^2 + \chi^4} + b \ln\left(\frac{b + \sqrt{b^2 + \chi^4}}{b + \sqrt{b^2 + 1}}    \right)                \right]
\label{eqkeralavarma2}
\end{equation}
with $\displaystyle{b^2 = \frac{\hat{h}_t}{3\hat{h}_q} + \alpha \frac{\hat{h}_t}{3\hat{h}_q} \frac{5}{8W^2 \chi^2}}$, $\alpha = [1 + \chi^2 - 5\chi^4 + 3\chi^6]/12$. Eq.~\ref{eqkeralavarma2} is an estimate of the coalescence stress (and not an upper-bound) as some approximations have been used to compute the plastic dissipation. Upper-bound estimates of the coalescence stress in absence of interface stresses can be found in \cite{thomasonnew2,morinthese} for alternative trial velocity fields. The purpose of the following section is to compute the second term of the RHS of Eq.~\ref{sumtwoterms}.

\subsection{Theoretical coalescence estimates}
\label{derivaxi}

Using Eqs.~\ref{criterion} and \ref{dissipationmisesnano3}, the coalescence stress accounting only for interface stresses can be written as:

\begin{equation}
\left(  \frac{\Sigma_{33}}{\sigma_0} \right)_{interface} = \frac{1}{D_{33} \sigma_0 \pi L^2 H} \int_{S_{int}} k_{int} d_{S,eq}^H\, dS 
\label{estimtheo}
\end{equation}
The non-zero components of the 3D strain rate tensor $\textbf{d}$ for the considered trial velocity field (Eq.~\ref{ve1}) are:
\begin{equation}
\left\{
\begin{aligned}
d_{rr} & = \frac{3D_{33} H}{4h} \left( -\frac{L^2}{r^2}-1   \right)\left( 1 - \frac{z^2}{h^2}   \right) \\
d_{\theta \theta} &= \frac{3D_{33} H}{4h} \left( \frac{L^2}{r^2}-1   \right)  \left( 1 - \frac{z^2}{h^2}   \right)\\
d_{zz}  &=  \frac{3D_{33} H}{2h}  \left( 1 - \frac{z^2}{h^2}   \right)  \\
d_{rz}  &=  -\frac{3D_{33}zH}{4h^3} \left( \frac{L^2}{r} - r   \right)\\
\end{aligned}
\right.
\label{KCcomponent}
\end{equation}
The void - matrix interface is composed of two parts $S_{int} = S_{int}^1 \cup S_{int}^2$ (Fig.~\ref{geometry}). On $S_{int}^2$, the velocity field corresponds to a rigid body motion, thus 2D strain rate tensor is equal to zero. On $S_{int}^1$, the 2D strain rate tensor is:

\begin{equation}
\textbf{d}^{S} = 
\left( \begin{array}{cc}
d_{\theta \theta } & d_{ \theta z}\\
d_{\theta z } & d_{zz} \\
  \end{array} \right) =
\left( \begin{array}{cc}
\displaystyle{\frac{3D_{33} H}{4h} \left( \frac{L^2}{R^2}-1   \right)  \left( 1 - \frac{z^2}{h^2}   \right)} &  0 \\
0  &  \displaystyle{\frac{3D_{33} H}{2h}  \left( 1 - \frac{z^2}{h^2}   \right)}\\ 
  \end{array} \right)
\end{equation}
Using the definition of the equivalent strain rate that appears in the integrand of the plastic dissipation:
\begin{equation}
d_{S,eq}^H  = \sqrt{\frac{2}{3}\textbf{d}^{\star}_{orth}:\mathbbm{\hat{h}}:\textbf{d}^{\star}_{orth}}
\label{2Dstrainhill}
\end{equation}
\begin{equation}
\textbf{d}^{\star}_{orth} = ^T\textbf{R}.\textbf{d}^{\star}.\textbf{R} = ^T\textbf{R} \left( \begin{array}{ccc}
-\mathrm{tr}(\textbf{d}^S) & 0 & 0 \\
0 & &  \\
0 & \multicolumn{2}{c}{\smash{\raisebox{.5\normalbaselineskip}{[$\textbf{d}^S$]}}}
  \end{array} \right)
\textbf{R}
\end{equation}
where \textbf{R} is the rotation matrix that goes from the cylindrical basis to the cartesian (orthotropy) basis:
\begin{equation}
\textbf{R} =  \left( \begin{array}{ccc}
\cos{\theta} & \sin{\theta} & 0 \\
-\sin{\theta} & \cos{\theta} & 0 \\
0 & 0 & 1
  \end{array} \right)
\end{equation}
allows to compute the plastic dissipation related to the interface (Eq.~\ref{estimtheo}):
\begin{equation}
\begin{aligned}
\left(  \frac{\Sigma_{33}}{\sigma_0} \right)_{interface} &= \frac{\Gamma}{\sqrt{3}\pi} \int_{\theta=0}^{2\pi} \sqrt{\hat{h}_c + 2(\hat{h}_t - \hat{h}_c)\sin^2{2\theta} - \hat{h}_d \cos{2\theta} \chi^2 + 3 \hat{h}_q \chi^4} d\theta \\
   &= \frac{\Gamma}{\sqrt{3}\pi}I_{\theta}(\mathbbm{\hat{h}},\chi)
\end{aligned}
\label{surstresshill}
\end{equation}
Combining Eq.~\ref{eqkeralavarma2} and Eq.~\ref{surstresshill}, we finally obtain the coalescence criterion for nanoporous anisotropic material:
\begin{equation}
\left(   \frac{\Sigma_{33}}{\sigma_0}\right)_c = \sqrt{\frac{6}{5} \hat{h}_q} \left[b \ln\frac{1}{\chi^2} + \sqrt{b^2+1} - \sqrt{b^2 + \chi^4} + b \ln\left(\frac{b + \sqrt{b^2 + \chi^4}}{b + \sqrt{b^2 + 1}}    \right)                \right] + \frac{\Gamma}{\sqrt{3}\pi}I_{\theta}(\mathbbm{\hat{h}},\chi) 
\label{critnanohill}
\end{equation}
For practical purposes, an analytical expression of the coalescence criterion might be preferred to the expression of Eq.~\ref{critnanohill} that requires performing the integration numerically. Restricting to transversely isotropic materials where loading axis $\underline{e}_3$ is aligned with the axis of material symmetry, $\hat{h}_t - \hat{h}_c=\hat{h}_d=0$, and the coalescence criterion is:
\begin{equation}
\left(   \frac{\Sigma_{33}}{\sigma_0}\right)_c = \sqrt{\frac{6}{5} \hat{h}_q} \left[b \ln\frac{1}{\chi^2} + \sqrt{b^2+1} - \sqrt{b^2 + \chi^4} + b \ln\left(\frac{b + \sqrt{b^2 + \chi^4}}{b + \sqrt{b^2 + 1}}    \right)                \right] + \frac{2\Gamma}{\sqrt{3}}\sqrt{\hat{h}_c + 3\hat{h}_q \chi^4}  
\label{critnanohilltaylor}
\end{equation}
Note that we consider for simplicity that the anisotropy of the matrix is the same as the one of the interface, but Eqs.~\ref{critnanohill} and \ref{critnanohilltaylor} also holds if the anisotropy is different (taking different values of $\hat{h}$ for the contribution of the interface and of the matrix). The second term in the RHS of Eqs.~\ref{critnanohill} and \ref{critnanohilltaylor} have been obtained without approximations when computing the plastic dissipation, for orthotropic and transversely isotropic materials where loading axis $\underline{e}_3$ is aligned with the axis of material symmetry, respectively. However, the first term in the RHS of Eq.~\ref{critnanohill} (same as in Eq.~\ref{critnanohilltaylor}) obtained in \cite{keralavarma} is only an approximation and not an upper-bound of the coalescence stress due to approximations made to compute the plastic dissipation. Therefore, Eqs.~\ref{critnanohill} and \ref{critnanohilltaylor} are only approximations of coalescence stress for cylindrical voids in axisymmetric loading conditions in the presence of interface stresses. Strict upper-bound estimates of coalescence stress without interface stresses have been proposed in \cite{thomasonnew2} for a von Mises material, and can replace the first term in the RHS of Eq.~\ref{critnanohill}, but this requires in principle to recompute the term due to interface stresses using the trial velocity field of \cite{thomasonnew2}. However, as shown in Appendix B, the result will be the same due the form of the trial velocity fields used in \cite{thomasonnew2,keralavarma}. Therefore, the first term in the RHS of Eqs.~\ref{critnanohill},\ref{critnanohilltaylor} can be replaced by strict upper-bounds of coalescence stress obtained in the absence of interface stresses (as the one given in \cite{thomasonnew2} for von Mises material) to get strict upper-bounds of coalescence stress with interface stresses.\\

\noindent
In the case of an isotropic matrix material and interface, the coalescence criterion obtained previously reduces to:

\begin{equation}
\left(   \frac{\Sigma_{33}}{\sigma_0}\right)_c = \sqrt{\frac{6}{5}} \left[b \ln\frac{1}{\chi^2} + \sqrt{b^2+1} - \sqrt{b^2 + \chi^4} + b \ln\left(\frac{b + \sqrt{b^2 + \chi^4}}{b + \sqrt{b^2 + 1}}    \right)                \right] + \frac{2\Gamma}{\sqrt{3}} \sqrt{1+3\chi^4} 
\label{critnanomises}
\end{equation}
with $\displaystyle{b^2 = \frac{1}{3} + \alpha \frac{1}{3} \frac{5}{8W^2 \chi^2}}$, $\alpha = [1 + \chi^2 - 5\chi^4 + 3\chi^6]/12$.

\subsection{Comparisons to numerical results}

Analytical estimates of coalescence stress for nanoporous material under axisymmetric loading conditions obtained in section~\ref{derivaxi} (Eqs.~\ref{critnanohill},\ref{critnanohilltaylor},\ref{critnanomises}) are assessed through comparisons to \textcolor{black}{supposedly exact (up to numerical errors)} results for spheroidal voids obtained by numerical limit analysis (Section~\ref{numana}). In all simulations presented hereafter, the matrix and the interface share the same anisotropy tensor $\mathbbm{\hat{h}}$ (Tab.~1).
\subsubsection{Elongated voids $W=3$}

\begin{figure}[H]
\centering
\subfigure[]{\includegraphics[height = 5cm]{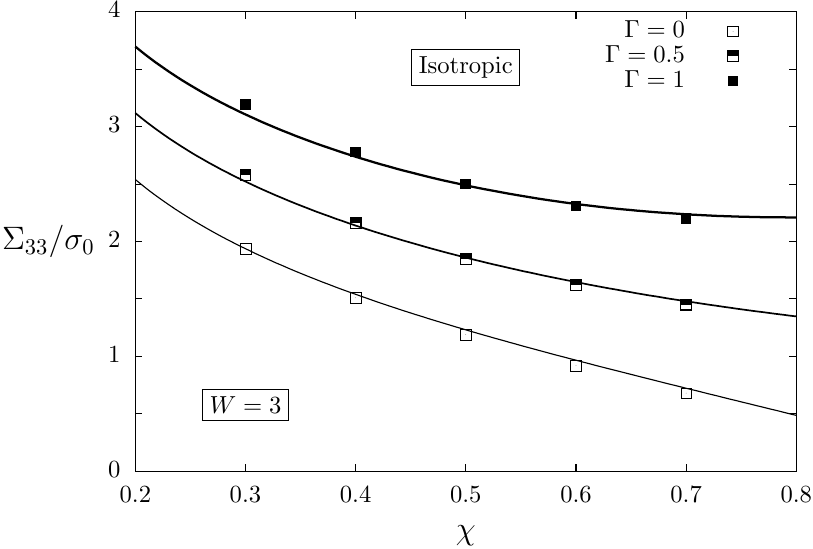}}
\subfigure[]{\includegraphics[height = 5cm]{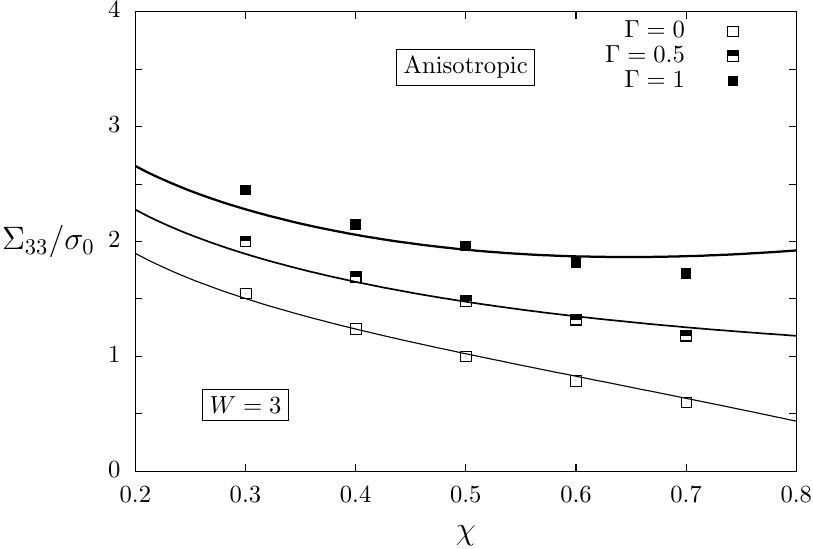}}
\caption{Limit-loads of elongated voids in cylindrical unit-cell. Comparison of the theoretical estimates to the results of numerical
limit analysis as a function of the inter-void ligament dimensionless length $\chi$, for different values of the dimensionless strength of the interface $\Gamma$.}
\label{fignum1}
\end{figure}

Comparisons are first made for elongated voids $W=3$, relevant for most practical situations, as initially flat or spherical voids are expected to become elongated before the onset of coalescence. Results are shown on Fig.~\ref{fignum1} for both materials described in Tab.~1, and for different values of the parameters $\chi$ (dimensionless length of the inter-void ligament) and $\Gamma$ (dimensionless strength of the interface).
A good agreement between analytical predictions and numerical results is observed for both isotropic and anisotropic materials considered and for a wide range of parameters ($\chi \in [0.3:0.7]$ and $\Gamma \in [0:1]$). Although additional simulations may be required to validate the analytical estimates for fully orthotropic materials, results presented in Fig.~\ref{fignum1} indicate that coalescence criteria derived in Section~\ref{derivaxi} may be used to describe nanoporous (transverse)-isotropic materials under axisymmetric loading conditions, for elongated voids.

\subsubsection{Spherical voids $W=1$}

\begin{figure}[H]
\centering
\subfigure[]{\includegraphics[height = 5cm]{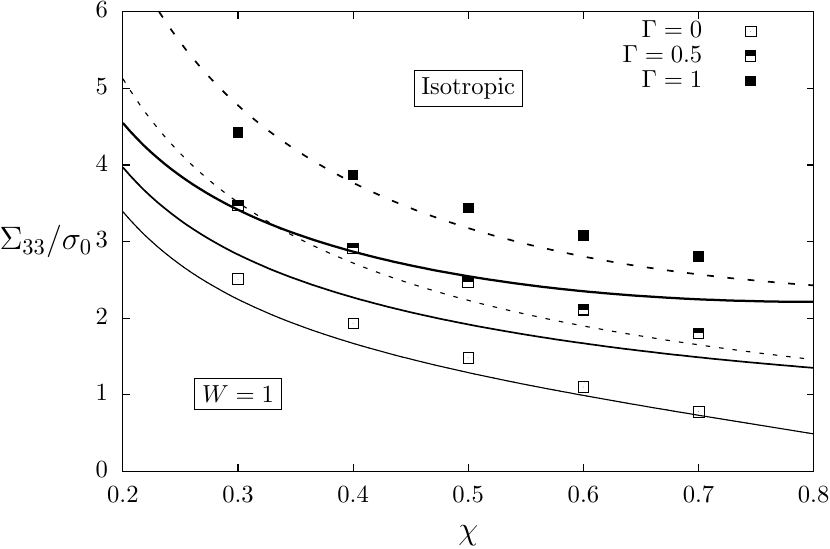}}
\subfigure[]{\includegraphics[height = 5cm]{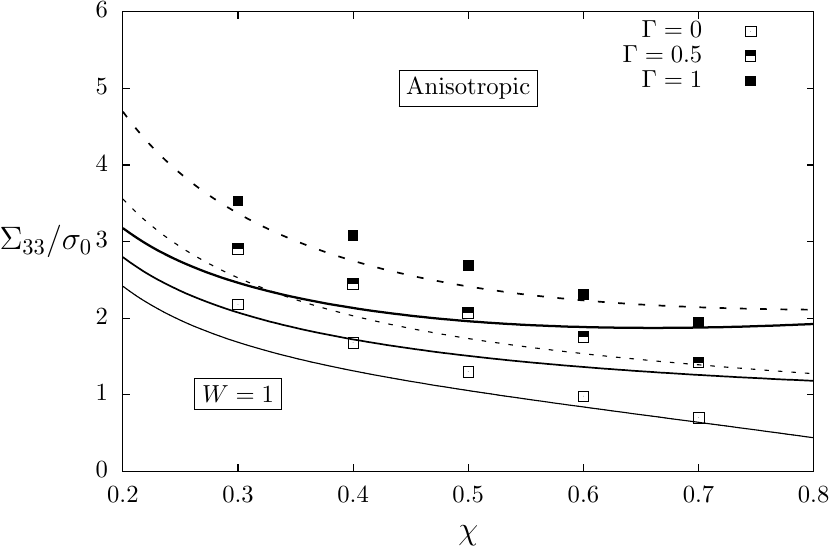}}
\caption{Limit-loads of equiaxed voids in cylindrical unit-cell. Comparison of the theoretical estimates to the results of numerical
limit analysis as a function of the inter-void ligament dimensionless length $\chi$, for different values of the dimensionless strength of the interface $\Gamma$. Solid lines correspond to criteria presented in Section~\ref{derivaxi}, dashed lines to the proposed phenomenological modification (Eq.~\ref{modifphenomeno}).}
\label{fignum2}
\end{figure}

For spherical voids ($W=1$), analytical predictions underestimate significantly coalescence stress obtained through numerical limit analysis, as shown by Fig.~\ref{fignum2} (solid lines), for both isotropic and anisotropic materials. While the agreement is rather good in the absence of interface stresses, coalescence criteria (Eqs.~\ref{critnanohilltaylor},~\ref{critnanomises}) fail to quantify the increase of coalescence stress with $\Gamma$. The origin of this discrepancy is assumed to come from the difference in void shapes between the one considered for numerical simulations and the one considered to get analytical predictions. While this difference has only a weak influence for elongated voids (where cylindrical and spheroidal shapes are rather close) (Fig.~\ref{fignum1}), a stronger effect is observed for spherical voids. One solution would be to perform analytical limit analysis considering spheroidal voids, but no trial velocity fields are currently available to describe accurately coalescence for such geometry. \textcolor{black}{Using cylindrical void shape in numerical simulations is neither a solution due to the complex plastic strain rate field (not described by the trial velocity field chosen in this study) arising because of the presence of the corner (as shown on Fig.~\ref{figstrainfield}).} A phenomenological modification is proposed here to tackle the case of spherical voids by considering an effective parameter:
\begin{equation}
\Gamma_{eff} = \mathcal{F}(W,\chi) \Gamma
\label{modifphenomeno}
\end{equation}
where the function $\mathcal{F}$ goes to zero for both $W \rightarrow +\infty$ and $\chi \rightarrow 1$, as the criteria of Eqs.~\ref{critnanohilltaylor},~\ref{critnanomises} are already in good agreement with numerical results in these limits. The choice $\mathcal{F}(W,\chi) = [0.5/W][\chi^{-1} - \chi]$ leads to the dashed lines in Fig.~\ref{fignum2}, in good agreement with numerical results, for both isotropic and anisotropic materials.

Comparisons made in this section indicate that the coalescence criteria derived in Section~\ref{derivaxi} are accurate to describe elongated spheroidal nanovoids in isotropic materials, and appear also to be accurate for anisotropic materials as shown for one particular case of transverse isotropy. For spherical voids, analytical predictions underestimate significantly numerical results, and a phenomenological modification has been proposed.\\

Coalescence criteria derived in Section~\ref{derivaxi} are only valid for axisymmetric loading conditions. They are extended in the next section for combined tension and shear loading conditions (as done in \cite{torki,keralavarma} in the limit $\Gamma = 0$).

\section{Combined tension and shear loading conditions}
\subsection{Theoretical coalescence estimates}

The previous coalescence criteria accounting for the presence of interface stresses are extended in this section in the presence of shear loading conditions. The starting point of the derivation is the approximate plastic dissipation proposed in \cite{keralavarma} (in absence of interface stresses):
\begin{equation}
\Pi^{app}_{\Gamma = 0}(\textbf{D}) = \sigma_0 \int_{\chi^2}^1 \left\{ D_{eq}^{S^2} + \frac{\overline{D}^2}{x^2}     \right\}^{1/2} dx
\end{equation}
where $\overline{D}$ can be viewed as an approximate axisymmetric equivalent strain rate, and $D_{eq}^{S^2} = \frac{2}{3}\textbf{D}^S:\mathbbm{\hat{h}}:\textbf{D}^S$, with $\textbf{D}^S$ is a macroscopic strain rate accounting only for shear components. Although only an approximation (see \cite{keralavarma}), the advantage of this form of plastic dissipation is that it is formally identical to the one obtained in the derivation of Gurson-like model, leading to well-known results for the form of the yield criterion \cite{revisited}. $\overline{D}$ was defined in \cite{keralavarma} so that to recover axisymmetric results when $D_{eq}^{S^2} \rightarrow 0$. Following the same approximation, we define $\overline{D}$ as:
\begin{equation}
\overline{D} \eqdef \left(\frac{\Sigma}{\sigma_0}   \right)_c\frac{D_{33}}{\ln(1/\chi^2)}
\label{definda}
\end{equation}
where $\left(\frac{\Sigma}{\sigma_0}   \right)_c$ is the axial coalescence stress obtained in Section~3 (Eq.~\ref{critnanohilltaylor} and Eq.~\ref{critnanomises}). Plastic dissipation accounting only for interface stresses in pure shear loading conditions ($\textbf{D}^S = D_{31}\underline{e}_3 \overset{S}{\otimes} \underline{e}_1  + D_{32}\underline{e}_3 \overset{S}{\otimes} \underline{e}_2$) is computed first using the trial shear velocity field given in Eq.~\ref{sheartrial}:
\begin{equation}
\begin{aligned}
\Pi_{\Gamma \neq 0}^{int}(\textbf{D}^S) &= \frac{1}{vol(\Omega)} \int_{S_{int}} k_{int} \sqrt{\frac{2}{3}\textbf{d}^{\star}_{orth}:\mathbbm{\hat{h}_{int}}:\textbf{d}^{\star}_{orth}}\, dS\\
                 &\leq \chi^2 \sigma_0 \Gamma \sqrt{\frac{2}{3}\textbf{D}^S:\mathbbm{\hat{h}_{int}}^S:\textbf{D}^S}
\end{aligned}
\end{equation} 
where Cauchy-Schwartz inequality has been used to get an upper-bound of the dissipation, and $\mathbbm{\hat{h}_{int}}^S$ is defined in Voigt-Mandel notation as:
\begin{equation}
\mathbbm{\hat{h}}^S \rightarrow
\left( \begin{array}{cccccc}
0 & 0 & 0 & 0 & 0 & 0\\
0 & 0 & 0 & 0 & 0 & 0\\
0 & 0 & 0 & 0 & 0 & 0\\
0     & 0     & 0     & \displaystyle{\frac{3\hat{h}_{44}+ \hat{h}_{55}}{4}}& 0 & 0\\
0     & 0     & 0     & 0 & \displaystyle{\frac{3\hat{h}_{55}+ \hat{h}_{44}}{4}} & 0\\
0     & 0     & 0     & 0 & 0 & 0
\end{array}
\right)
\label{hshear}
\end{equation}
The approximate plastic dissipation accounting for both interface stresses and matrix in the case of pure shear loading is:
\begin{equation}
\Pi_{\Gamma \neq 0}^{app}(\textbf{D}^S) = \sigma_0  \left[ (1-\chi^2)\sqrt{\frac{2}{3}\textbf{D}^S:\mathbbm{\hat{h}_{mat}}:\textbf{D}^S} + \chi^2 \Gamma \sqrt{\frac{2}{3}\textbf{D}^S:\mathbbm{\hat{h}_{int}}^S:\textbf{D}^S} \right]
\end{equation}
where $\mathbbm{\hat{h}_{mat}}$ is the anisotropy tensor of the matrix material, and $\mathbbm{\hat{h}_{int}}^S$ is defined in Eq.~\ref{hshear} using the parameters of the anisotropy tensor of the interface. In order to get analytical estimate of the yield criterion, we now restrict to the case where both matrix material and interface are transverse isotropic with respect to the $\underline{e}_3$ axis. The plastic dissipation can be rewritten as:
\begin{equation}
\Pi_{\Gamma \neq 0}^{app}(\textbf{D}^S) = \sigma_0  (1-\chi^2) \left[1 + \sqrt{\frac{\hat{h}_{44}^{int}}{\hat{h}_{44}^{mat}}}\frac{\chi^2 \Gamma}{(1-\chi^2)}      \right]   \sqrt{\frac{2}{3}\textbf{D}^S:\mathbbm{\hat{h}_{mat}}:\textbf{D}^S}
\label{shearhh}
 \end{equation}
Finally, the global macroscopic dissipation is approximated by the formula:
\begin{equation}
\Pi^{app}_{\Gamma \neq 0}(\textbf{D}) = \sigma_0 \int_{\chi^2}^1 \left\{ \overline{D}_{S}^{2} + \frac{\overline{D}^2}{x^2}     \right\}^{1/2} dx
\label{eqdissipa}
\end{equation}
where $\overline{D}_{S}$ is defined so as to recover Eq.~\ref{shearhh} for pure shear:
\begin{equation}
\begin{aligned}
\overline{D}_{S}^2 &= \left(1 +\sqrt{\frac{\hat{h}_{44}^{int}}{\hat{h}_{44}^{mat}}}\frac{\chi^2 \Gamma}{(1-\chi^2)}      \right)^2  \frac{2}{3}\textbf{D}^S:\mathbbm{\hat{h}_{mat}}:\textbf{D}^S \\
                  &= \mathcal{G}(\chi,\Gamma,\mathbbm{\hat{h}_{mat}},\mathbbm{\hat{h}_{int}})\frac{2}{3}\textbf{D}^S:\mathbbm{\hat{h}_{mat}}:\textbf{D}^S 
\end{aligned}
\label{definds}
\end{equation}
Following Gurson's lemma \cite{revisited} using the approximate plastic dissipation of Eq.~\ref{eqdissipa} and definitions of Eqs.~\ref{definda},~\ref{definds}, the close-formed equation of the coalescence criterion is:
\begin{equation}
\phi(\bm{\Sigma},\chi,W,\Gamma,\mathbbm{\hat{h}_{mat}},\mathbbm{\hat{h}_{int}}) = \frac{3}{2\mathcal{G}} \frac{\bm{\Sigma}^S:\mathbbm{p_{mat}}:\bm{\Sigma}^S}{\sigma_0^2} + 2 \chi^2 \cosh{\left(   \frac{\Sigma_{33}}{\Sigma_c} \ln{\frac{1}{\chi^2}}     \right)} -1 - \chi^4 = 0
\label{critfinalshear}
\end{equation}
where $\bm{\Sigma}^S = \Sigma_{31}\underline{e}_3 \overset{S}{\otimes} \underline{e}_1 + \Sigma_{32} \underline{e}_3 \overset{S}{\otimes} \underline{e}_2$, and $\mathcal{G}$ and $\Sigma_c$ are defined by Eqs.~\ref{definds},~\ref{critnanohilltaylor}, respectively. Eq.~\ref{critfinalshear} has been derived through various approximations for transversely isotropic material (and interface) when the loading axis $\underline{e}_3$  is aligned with the axis of material symmetry, thus requiring validation through comparisons to numerical limit analysis. 
\subsection{Comparisons to numerical results}

Coalescence criterion obtained in the previous section is compared to the results of numerical limit analysis for elongated $W=3$ and spherical $W=1$ voids, and different values of the parameters $\chi$ and $\Gamma$, considering isotropic material and interface. For spherical voids, the phenomenological modification (Eq.~\ref{modifphenomeno}) for the coalescence stress in axisymmetric loading conditions (Eq.~\ref{critnanohilltaylor}) is taken into account.
\begin{figure}[H]
\centering
\subfigure[]{\includegraphics[height = 5cm]{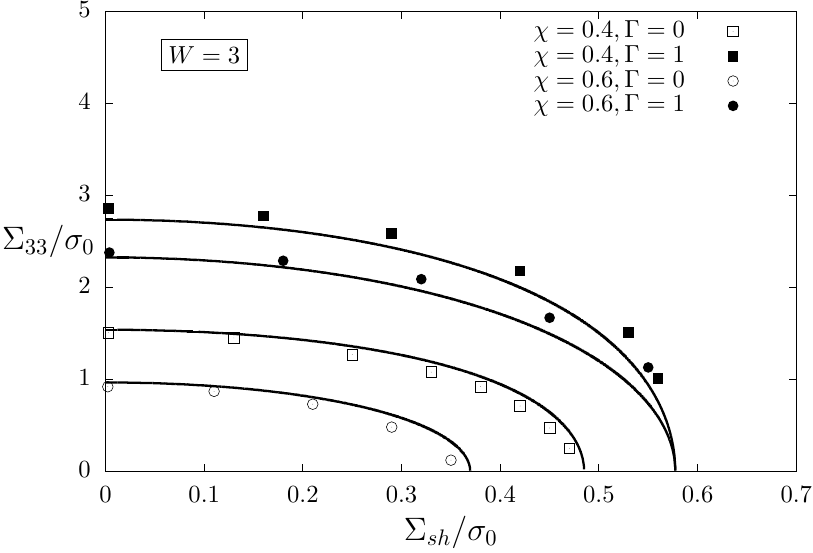}}
\subfigure[]{\includegraphics[height = 5cm]{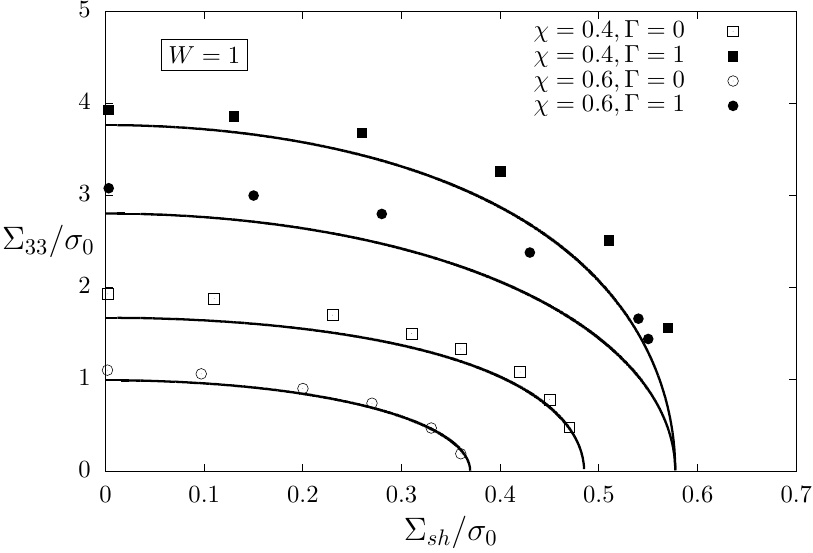}}
\caption{Yield locus of (nano)-voids in cylindrical unit-cell for isotropic matrix / interface. Comparison of the coalescence criterion (Eq.~\ref{critfinalshear}) to the results of numerical limit analysis, for different values of  the inter-void ligament dimensionless length $\chi$ and dimensionless strength of the interface $\Gamma$. ($\Sigma_{sh}^2 = \Sigma_{13}^2 +  \Sigma_{23}^2$) }
\label{fignum3}
\end{figure}
Analytical criterion (Eq.~\ref{critfinalshear}) is found to be in good agreement with numerical results for the yield locus of spheroidal (nano)-voids, for the case of isotropic matrix and interface (Fig.~\ref{fignum3}). In particular with interface stresses, the model is able to reproduce a peculiar feature due to the chosen parameters which is that the macroscopic yield stress in pure shear loading conditions is the same for two different values of intervoid ligament $\chi$. \textcolor{black}{Note that the numerical results presented in Fig.~\ref{fignum3} are slightly different from the ones reported in \cite{torki2017} due to different void shapes (cylindrical for \cite{torki2017} \textit{vs.} spheroidal in this study).}

\section{Discussion and Conclusions}
\label{conclusion}
\begin{figure}[H]
\centering
\subfigure[]{\includegraphics[height = 5cm]{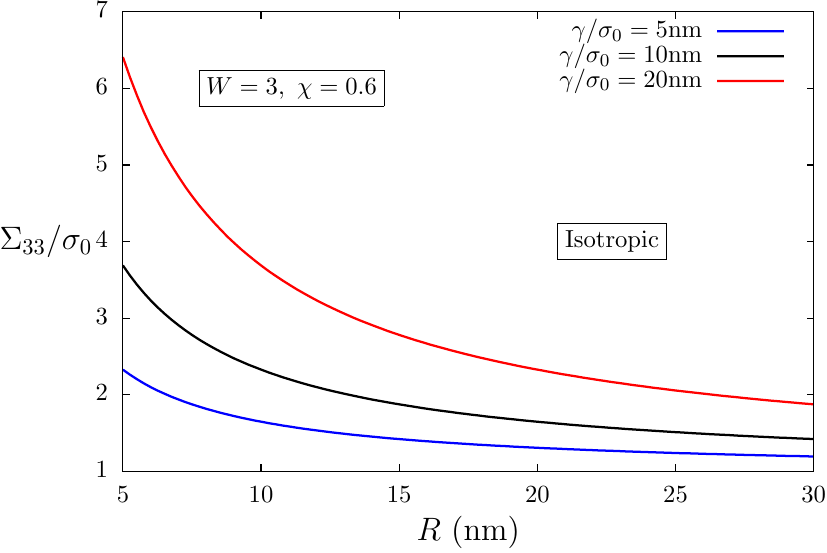}}
\subfigure[]{\includegraphics[height = 5cm]{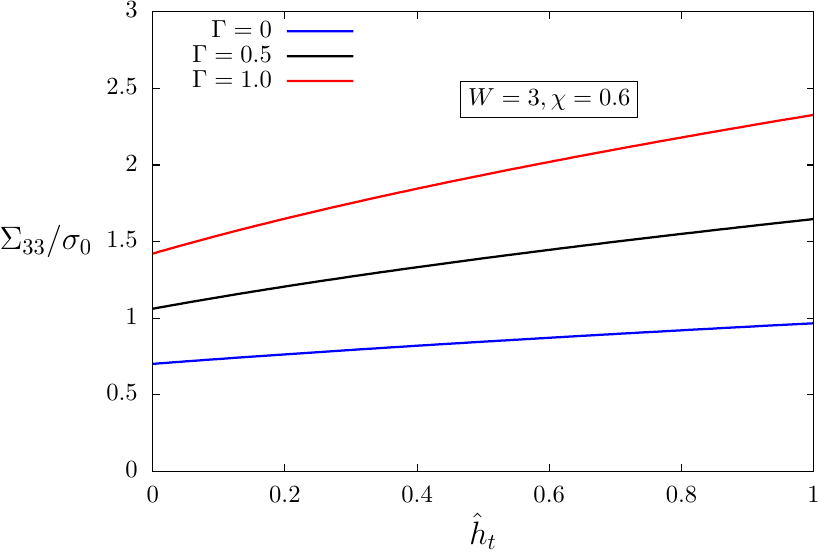}}
\caption{Coalescence stress in axisymmetric loading conditions from Eqs.~\ref{critnanohilltaylor},~\ref{critnanomises}: (a) for isotropic matrix and interface as a function of the size of the void $R$; (b) for transversely isotropic matrix and interface (with $\hat{h}_q=1$, $\hat{h}_d=0$), as a function of $\hat{h}_t = \hat{h}_c$.}
\label{figconclusion}
\end{figure}

Nanovoids are observed in materials of practical importance such as metal alloys used in nuclear power plants as a consequence of irradiation. Growth and coalescence of nano-voids might differ from their microscopic counterparts for several reasons. On one hand, the scarcity of dislocations at the scale of the void entails void deformation unless high applied stress allows nucleation of dislocations from the void surface, and/or because the presence of GNDs induce additional strain-hardening. On the other hand, energetic considerations indicate that interface energy at void/matrix interface should be accounted for when the void size $R$ is small compared to the typical lengthscale $\gamma/\sigma_0$, where $\gamma$ is the surface energy and $\sigma_0$ is the strength of the matrix. These mechanisms lead to size effects regarding void deformation under mechanical loading. In addition, the matrix surrounding nanovoids in crystalline materials is anisotropic, due to slip systems activity. Following the limit analysis framework proposed and applied to obtain homogenized models for nanoporous materials in the growth regime in \cite{dormieux2010,monchiet2013}, analytical coalescence criterion accounting for interface stresses / surface energy is provided for axisymmetric loading conditions and orthotropic plastic matrix material, incidentally extending the isotropic modeling of interface stresses proposed in  \cite{dormieux2010} to anisotropy. Coalescence criterion has also been provided for combined tension and shear loading conditions. Some key results are shown on Fig.~\ref{figconclusion}, where the effects of void size (Fig.~\ref{figconclusion}a) and anisotropy of the matrix (Fig.~\ref{figconclusion}b) on coalescence stress are quantified. \textcolor{black}{For void size below the typical lengthscale $\gamma/\sigma_0$, significant strengthening of nanoporous material is observed due to the interfacial energy requires for void to deform, for both the growth regime \cite{monchiet2013} and in the coalescence regime (Fig.~\ref{figconclusion}). For a given porosity, nanoporous ductile materials are thus harder than their microscopic counterparts}.

The proposed analytical coalescence criteria have been \textcolor{black}{successfully} validated \textcolor{black}{(for specific conditions)} through comparisons to numerical results \textcolor{black}{for elongated spheroidal voids, for axisymmetric and combined tension / shear loading conditions, and for isotropic and orthotropic materials. However, a phenomenological modification is necessary for the criteria to be used for spherical voids. Refined trial velocity fields are needed to tackle this issue.} \textcolor{black}{The case of flat or penny-shaped voids ($W \ll 1$) was not considered in this study as nanovoid coalescence was not observed in numerical simulations under axisymmetric loading conditions. Occurence of penny-shaped nanovoids coalescence under intense shearing, as well as the dependence of coalescence stress on void shape as illustrated in Fig.~\ref{figstrainfield}, deserve attention and should be the object of a future study.}

Following the hybrid methodology detailed in \cite{benzergaleblond}, considering both yield criteria for the growth phase \cite{monchiet2013} and for the coalescence phase lead to a complete homogenized model of nanoporous materials, adding evolution laws for state variables. \textcolor{black}{Such hybrid model will depend on the value of the parameter $\Gamma$ that will decrease due to the evolution of the void radius, for example in the growth regime as shown in \cite{morin2015}, ensuring a seamless transition from nano- ($\Gamma \neq 0$) to micro-porous material ($\Gamma \rightarrow 0$).} Such work on the hybrid model in underway and will be presented elsewhere. 

While the modeling of nanoporous materials used in this study accounts for surface energy of the void-matrix interface, the proposed criterion (and the ones described in \cite{dormieux2010,monchiet2013}) could be used in broader situations, assuming that the 2D yield stress of the interface is not (only) the surface tension of the interface, but is taken as an effective parameter describing additional dissipation at the void surface, like for example hardening due to GNDs. This effective parameter should ultimately be adjusted on experimental data, which are still to a great extent lacking in the literature.  \\

\noindent
\textbf{Acknowledgments}\\

The authors would like to thank B. Tanguy for the careful reading of the manuscript. Discussions with D. Kondo and J.B. Leblond are also acknowledged. This project has received funding from the Euratom research and training programme 2014-2018 under Grant Agreement N$^{\circ}$661913. This work reflects only the authors' view and the Commission is not responsible for any use that may be made of the information it contains.

\section{Appendix A}

For completeness, Voigt-Mandel notations used in this study and definitions of plastic anisotropy tensors are given in this section. The reader is referred to \cite{benzergaanisotrope} for additional details. In Voigt-Mandel notations, and for an orthonormal basis $\{\underline{e}_1,\underline{e}_2,\underline{e}_3\}$, stress and strain rate second order tensors are represented as vectors such that:

\begin{equation}
\bm{\sigma} \rightarrow
\left( \begin{array}{c}
\sigma_{11}\\
\sigma_{22}\\
\sigma_{33}\\
\sqrt{2}\sigma_{23}\\
\sqrt{2}\sigma_{13}\\
\sqrt{2}\sigma_{12}
\end{array}
\right) 
\ \ \ \ \ 
\textbf{d} \rightarrow
\left( \begin{array}{c}
d_{11}\\
d_{22}\\
d_{33}\\
\sqrt{2}d_{23}\\
\sqrt{2}d_{13}\\
\sqrt{2}d_{12}\\
\end{array}
\right)
\end{equation}
For orthotropic materials obeying Hill's criterion \cite{hill} ($\sigma_{eq}^{H}= \sqrt{\frac{3}{2}\bm{\sigma}:\mathbbm{p}:\bm{\sigma}}$ and $d_{eq}^H = \sqrt{\frac{2}{3}\textbf{d}:\mathbbm{\hat{h}}:\textbf{d}}$), fourth-order tensor $\mathbbm{p}$ and $\hat{\mathbbm{h}}$ defining the plastic anisotropy of the material  are represented as matrix:
\begin{equation}
\mathbbm{p} \rightarrow
\left( \begin{array}{cccccc}
p_{11} & p_{12} & p_{13} & 0 & 0 & 0\\
p_{12} & p_{22} & p_{23} & 0 & 0 & 0\\
p_{13} & p_{23} & p_{33} & 0 & 0 & 0\\
0     & 0     & 0     & p_{44} & 0 & 0\\
0     & 0     & 0     & 0 & p_{55} & 0\\
0     & 0     & 0     & 0 & 0 & p_{66}
\end{array}
\right) 
\ \ \ \ \ 
\mathbbm{\hat{h}} \rightarrow
\left( \begin{array}{cccccc}
\hat{h}_{11} & \hat{h}_{12} & \hat{h}_{13} & 0 & 0 & 0\\
\hat{h}_{12} & \hat{h}_{22} & \hat{h}_{23} & 0 & 0 & 0\\
\hat{h}_{13} & \hat{h}_{23} & \hat{h}_{33} & 0 & 0 & 0\\
0     & 0     & 0     & \hat{h}_{44} & 0 & 0\\
0     & 0     & 0     & 0 & \hat{h}_{55} & 0\\
0     & 0     & 0     & 0 & 0 & \hat{h}_{66}
\end{array}
\right)
\end{equation}
With these notations, equivalent stress and strain rate can be computed by matrix product. Components of $\mathbbm{p}$ and $\mathbbm{\hat{h}}$ are related through the formulas:
\begin{equation}
\mathbbm{\hat{p}} = \mathbbm{J}:\mathbbm{\hat{h}}:\mathbbm{J}
\end{equation}
\begin{equation}
\mathbbm{\hat{p}}:\mathbbm{p} =\mathbbm{p}: \mathbbm{\hat{p}} = \mathbbm{J}
\label{defdeq}
\end{equation}
where $\mathbbm{\hat{p}}$ is an alternative fourth-order tensor to compute the equivalent strain rate ($d_{eq}^H = \sqrt{\frac{2}{3}\textbf{d}:\mathbbm{\hat{p}}:\textbf{d}}$), and $\mathbbm{J} = \mathbbm{I} - (1/3) \textbf{I}\otimes \textbf{I}$ is the projection operator, which reads in Voigt-Mandel notation:

\begin{equation}
J \rightarrow
\left( \begin{array}{cccccc}
2/3 & -1/3 & -1/3 & 0 & 0 & 0\\
-1/3 & 2/3 & -1/3 & 0 & 0 & 0\\
-1/3 & -1/3 & 2/3 & 0 & 0 & 0\\
0     & 0     & 0     & 1 & 0 & 0\\
0     & 0     & 0     & 0 & 1 & 0\\
0     & 0     & 0     & 0 & 0 & 1
\end{array}
\right) 
\end{equation} 
Eq.~\ref{defdeq} comes from the definition of the equivalent strain rate and from the principle of work equivalence $\sigma_{eq}^H d_{eq}^H = \bm{\sigma}:\textbf{d}$, and can be rewritten as \cite{morinthese}:

\begin{equation}
\left\{
\begin{aligned}
\hat{p}_{11} &= -\hat{p}_{12} - \hat{p}_{13}\\
\hat{p}_{22} &= -\hat{p}_{12} - \hat{p}_{23}\\
\hat{p}_{33} &= -\hat{p}_{13} - \hat{p}_{23}\\
\hat{p}_{44} &= \frac{1}{p_{44}}\\
\hat{p}_{55} &= \frac{1}{p_{55}}\\
\hat{p}_{66} &= \frac{1}{p_{66}}\\
\hat{p}_{12} &= -\frac{1}{9}\frac{p_{12} - 2p_{13} - 2p_{23}}{p_{12}p_{13}+p_{12}p_{23} + p_{13}p_{23}}\\
\hat{p}_{13} &= -\frac{1}{9}\frac{p_{13} - 2p_{12} - 2p_{23}}{p_{12}p_{13}+p_{12}p_{23} + p_{13}p_{23}}\\
\hat{p}_{23} &= -\frac{1}{9}\frac{p_{23} - 2p_{12} - 2p_{13}}{p_{12}p_{13}+p_{12}p_{23} + p_{13}p_{23}}\\
\end{aligned}
\right.
\label{relationAB}
\end{equation}
For an isotropic material (obeying von Mises criterion), fourth-order tensor $\mathbbm{p}$ and $\mathbbm{\hat{h}}$ are represented in Voigt-Mandel notation as:
\begin{equation}
\mathbbm{p} \rightarrow
\left( \begin{array}{cccccc}
2/3 & -1/3 & -1/3 & 0 & 0 & 0\\
-1/3 & 2/3 & -1/3 & 0 & 0 & 0\\
-1/3 & -1/3 & 2/3 & 0 & 0 & 0\\
0     & 0     & 0     & 1 & 0 & 0\\
0     & 0     & 0     & 0 & 1 & 0\\
0     & 0     & 0     & 0 & 0 & 1
\end{array}
\right) 
\ \ \ \ \ 
\mathbbm{\hat{h}} \rightarrow
\left( \begin{array}{cccccc}
1 & 0 & 0 & 0 & 0 & 0\\
0 & 1 & 0 & 0 & 0 & 0\\
0 & 0 & 1 & 0 & 0 & 0\\
0     & 0     & 0     & 1 & 0 & 0\\
0     & 0     & 0     & 0 & 1 & 0\\
0     & 0     & 0     & 0 & 0 & 1
\end{array}
\right)
\end{equation}

\section{Appendix B}
Trial velocity fields used in \cite{thomasonnew2}(continuous one) and in \cite{keralavarma} can be written as:
\begin{equation}
\left\{
\begin{aligned}
v_r^i(r,z) & = f^i(z)g(r) \\
v_z^{i}(r,z) &= p^i(z)\\
\end{aligned}
\right. 
\label{trialvelocityfieldform}
\end{equation}
where the superscript $i$ refers to Keralavarma and Chockalingham or Morin-Leblond-Benzerga choices, but same radial dependence (through the function $g(r)$) is taken in both cases. $f^i(z)$ and $p^i(z)$ must satisfy the following conditions coming from boundary conditions and the requirement of incompressiblity:
\begin{equation}
\left\{
\begin{aligned}
f^i\left(g' + \frac{g}{r}\right) + (p^i)' &= 0 \\
p^i(0) &= 0\\
p^i(h) &= HD_{33}\\
\end{aligned}
\right. 
\label{trialvelocityfieldform2}
\end{equation}
where $(\ )^'$ stands as the derivative with respect to the only variable. Computing the 2D equivalent strain rate of the interface (Eq.~\ref{2Dstrainhill}) with a velocity field satisfying Eqs.~\ref{trialvelocityfieldform},~\ref{trialvelocityfieldform2} on the interface $S_{int}^1$ (Fig.~\ref{geometry}) leads to $d_{S,eq}^H = f^i(z) \mathcal{D}(R,\theta,\mathbbm{\hat{h}})$. Upon integration, the plastic dissipation is 
\begin{equation}
\begin{aligned}
\Pi &= \frac{1}{vol{\Omega}} \int_{S_{int}^1} k_{int} d_{S,eq}^H \,dS \\
    &= \frac{k_{int}}{vol{\Omega}} \int_0^h f^i(z)dz \int_{0}^{2\pi} \mathcal{D}(R,\theta,\mathbbm{\hat{h}}) Rd\theta \\
    &= \frac{k_{int}}{vol{\Omega}} \int_0^h -\frac{(p^i(z))'}{\displaystyle{g'(R) + \frac{g(R)}{R}}}dz \int_{0}^{2\pi} \mathcal{D}(R,\theta,\mathbbm{\hat{h}}) Rd\theta \\
    &= -\frac{k_{int}}{vol{\Omega}}   \frac{HD_{33}}{\displaystyle{g'(R) + \frac{g(R)}{R}}} \int_{0}^{2\pi} \mathcal{D}(R,\theta,\mathbbm{\hat{h}}) Rd\theta\\
\end{aligned}
\end{equation}
which is independent of the choice of the functions $f^i$ and $p^i$. For any trial velocity field satisfying conditions of Eqs.~~\ref{trialvelocityfieldform},~\ref{trialvelocityfieldform2}, the plastic dissipation related to interface stresses on the interface $S_{int}^1$ is thus the same, and equal to the one computed in section~\ref{derivaxi}.

\bibliographystyle{elsarticle-num.bst}
\bibliography{spebib2}

\end{document}